\documentclass[%
 superscriptaddress,
 amsmath,amssymb,
 aps, pre
]{revtex4}

\usepackage[usenames,dvipsnames]{xcolor}
\usepackage{amsthm}
\usepackage{amssymb}

\newtheorem{theorem}{Theorem}[section]

\usepackage{graphicx}
\usepackage{dcolumn}
\usepackage{bm}
\usepackage[T2A]{fontenc}
\usepackage[utf8]{inputenc}
\usepackage{comment}
\usepackage{bigstrut}
\usepackage{float}

\usepackage[unicode, colorlinks, urlcolor=blue, linkcolor=blue, citecolor=blue]{hyperref} 
\begin{document}

\title{Systematic derivation of angular--averaged Ewald potential}

\author{G. S. Demyanov}
\affiliation{Joint Institute for High Temperatures, Izhorskaya 13 Bldg 2, Moscow 125412, Russia}
\affiliation{Moscow Institute of Physics and Technology, Institutskiy Pereulok 9, Dolgoprudny, Moscow Region, 141701, Russia}
\author{P. R. Levashov}
\affiliation{Joint Institute for High Temperatures, Izhorskaya 13 Bldg 2, Moscow 125412, Russia}
\affiliation{Moscow Institute of Physics and Technology, Institutskiy Pereulok 9, Dolgoprudny, Moscow Region, 141701, Russia}

\date{\today}

\begin{abstract}
In this work we provide a step by step derivation of an angular--averaged Ewald potential suitable for numerical simulations of disordered Coulomb systems. The potential was first introduced by E.\,Yakub and C.\,Ronchi without a clear derivation. Two methods are used to find the coefficients of the series expansion of the potential: based on the Euler--Maclaurin and Poisson summation formulas. The expressions for each coefficient is represented as a finite series containing derivatives of Jacobi theta functions. We also demonstrate the formal equivalence of the Poisson and Euler--Maclaurin summation formulas in the three-dimensional case. The effectiveness of the angular--averaged Ewald potential is shown by the example of calculating the Madelung constant for a number of crystal lattices. 
\end{abstract}

\maketitle

\section{Introduction}
The Coulomb potential plays a fundamental role in numerous theoretical and applied problems \cite{Kalman:SCCS:1998}. Any system of charged particles is characterized by long--range electrostatic interaction which originates the main problem in the mathematical description of such systems. Electrostatic energy of an infinite electroneutral system of charged particles is a \emph{conditionally} convergent series; its sum depends on the summation order \cite{UnderstandingMolecularSimulation}. The solution to this problem for systems with a translational symmetry was proposed by Ewald \cite{Ewald:1921}. By addition and subtraction of normally--distributed screening charges the original sum may be transformed into two rapidly converging sums. The correctness of the Ewald's summation technique is justified experimentally. 

However this subtle approach can't be directly applied to disordered Coulomb systems. By the term ``disordered system'' we mean a system in which there are only small correlations (or no correlations at all) between the positions of the particles. 
In other words, the pair correlation function of ions positions demonstrates only short--range order and becomes constant rather quickly. As the Coulomb potential is isotropic we assume the isotropy of a whole Coulomb system. To simulate a disordered system of charged particles one considers a significantly large (mostly cubic) computational cell; we assume that the number of particles $N$ in the cell is large ($\ln N \gg 1$). Periodic boundary conditions are imposed on the cell so that an infinite anisotropic system with translation symmetry forms. 
By increasing the number of particles in the cell it is possible to find the thermodynamic limit, i.e. energy per particle at $N\to\infty$.

Isotropic potentials are widely used in atomistic modeling of liquid and plasma media. 
The Ewald's technique defines an anisotropic potential being artificial and reduntant for disordered systems of particles (ionic liquids, plasma). 
Nevertheless, in atomistic simulations of plasma the Ewald's summation is widely used despite the fact that the computational effort scales as $O(N^{3/2})$ \cite{Baus:PR:1980}. There are also other approaches concerning the problem of long--range potentials, including fast multipole \cite{Greengard:JCP:1987}, particle-mesh-based methods \cite{Eastwood:JCP:1974} and smooth particle mesh Ewald method \cite{Essmann:1995}; however, they become efficient only for sufficiently large systems (at least $10^5$ particles). 

In 2003 E.\,Yakub and C.\,Ronchi \cite{Yakub:2003} proposed an angular--averaging technique for the Ewald potential and showed numerically the consistency of the new potential. Their result was presented as a power series depending on the distance between particles and it was stated that all the coefficients of the series except for the first two were equal to zero. The potential was then used in many computational works due to its obvious efficiency \cite{Yakub:2005,Yakub:JPA:2006,Jha:2010,Filinov:PRE:2020, Yakub:2007, Fukuda:2011, Fukuda:2012, Guerrero:2011, Fukuda:2013, Guo:2011, Lytle:2016, Nikitin:2020, KAMIYA201326}. However, no systematic mathematical derivation of the angular--averaged Ewald potential was published in the literature. Therefore, the question remained whether the new potential is approximate?
Thus, the main purpose of our work is to provide a step by step derivation of the fundamental formulas in \cite{Yakub:2003} from the original Ewald potential and analyze their effectiveness.

We investigate the coefficients of the power series for the angular--averaged Ewald potential including their dependence on the smearing parameter. We show that all the coefficients except for the first two tend to zero in the case of point charges. Two methods are used to find the coefficients of the series expansion of the potential: based on the Euler--Maclaurin and Poisson summation formulas. The expressions for each coefficient is represented as a finite series containing derivatives of Jacobi theta functions. We also demonstrate the formal equivalence of the Poisson and Euler--Maclaurin summation formulas in the three--dimensional case. The physical meaning of the potential is discussed including the fulfillment of the elecroneutrality condition. Finally, we demonstrate the convergence of the Madelung constant for a number of crystal lattices using the direct summation with the angular--averaged Ewald potential for up to $2\times 10^7$ particles. 

The article is organized as follows. Section~\ref{sec:problem} contains the description of 
the series summation problem by averaging the Ewald potential over all directions.
In Section~\ref{sec:summation} we sum the series using the Poisson formula and obtain the final expression for the averaged potential \eqref{eq:averagedPot} (or Eq. (6) in the original work \cite{Yakub:2003}).
In Section~\ref{sec:analysis} we analyze the obtained averaged potential and find out its physical meaning.
In Section~\ref{sec:appl} we present the calculation of Madelung constants for ordered systems using the averaged potential, analyze their convergence rate, and examine the performance of this calculation method. We summarize our study in Section~\ref{Sec:conclusions}. 

\section{Summation problem}
\label{sec:problem}
Consider a cubic cell of a volume $L^3$ which contains $N$ point particles. Each $i$-th particle has a charge $Q_i$ and position $\textbf{r}_i$ in the cube. The system is electroneutral:
\begin{equation}
\label{eq:electroneutral}
\sum_{i = 1}^NQ_i = 0.
\end{equation}
Periodic boundary conditions are assumed, so the cell repeats itself in three mutually perpendicular directions. It means that a particle with a position $\textbf{r}_i$ in the cell has infinite number of images with positions $\textbf{r}_i + \textbf{n} L$. Here, $\textbf{n}$ is an integer vector $\textbf{n}~=~(n_x, n_y, n_z)$, $n_x, n_y, n_z \in \mathbb{Z}$.

According to Coulomb's law, the total potential energy $E$ of such an infinite system is (we use Gaussian units):
\begin{equation}
E = \cfrac{1}{2}\,\sideset{}{'}\sum_{\textbf{n}}\sum_{i=1}^N\sum\limits_{j=1}^N\cfrac{Q_iQ_j}{|\textbf{r}_i - \textbf{r}_j + L\textbf{n}|}.
\end{equation}
The summation is performed over all integer vectors $\textbf{n}$; the prime means that the terms with $\textbf{n}=\textbf{0}$ are omitted if $i = j$. So a particle $i$ interacts with all its replica images, but not with itself. This sum is \emph{conditionally convergent}; thus, to obtain the correct answer one has to use a special Ewald summation technique \cite{Ewald:1921}. The main idea is to add and subtract a normally distributed screening charge  with a standard deviation $\sqrt{L^2/(2\delta^2)}$ \cite[p. 294]{UnderstandingMolecularSimulation}; $\delta>0$ is a dimensionless parameter. Below we assume the dependence of all values on $\delta$; point charges correspond to the case $\delta\to\infty$. Ewald summation  procedure results in \cite[p. 346]{rapaport_2004}:
\begin{equation}
	\label{eq:ewaldexact}
	E = \phi_1^{\text{Ex}}
	\sum_{i = 1}^NQ_i^2+\cfrac{1}{2}\sum_{i = 1}^N\sum_{\substack{j = 1\\
			i\neq j}}^N Q_iQ_j\phi^{\text{Ex}}_2(\textbf{r}_{ij}),	
\end{equation}
\begin{equation}
\phi_1^{\text{Ex}} = \cfrac{1}{2L}\left[
\sum_{\textbf{n}\neq\textbf{0}}\left(\cfrac{\mathrm{erfc}(\delta n)}{n} + \cfrac{1}{\pi n^2}\exp\left(-\cfrac{\pi^2n^2}{\delta^2}\right)\right) - \cfrac{2\delta}{\sqrt{\pi}}
\right],
\end{equation}
\begin{equation}
\phi^{\text{Ex}}_2(\textbf{r}_{ij}) = \cfrac{1}{L}\left[
\sum_{\textbf{n}}\cfrac{\mathrm{erfc}\left(\delta|\textbf{r}_{ij}/L + \textbf{n}|\right)}{|\textbf{r}_{ij}/L + \textbf{n}|} +
\cfrac{1}{\pi }\sum_{\textbf{n}\neq\textbf{0}}\cfrac{1}{n^2}\exp\left(-\cfrac{\pi^2n^2}{\delta^2}\right)\cos\left(\cfrac{2\pi}{L}\,\textbf{n}\cdot \textbf{r}_{ij}\right)
\right],
\end{equation}
where $\mathrm{erfc}(x)$ is the complementary error function. The summation $\sum_{\textbf{n}\neq\textbf{0}}$ means that the term $\textbf{n}=(0,0,0)$ is omitted for all $i, j$. Here, $n = |\textbf{n}|$, $\textbf{r}_{ij} = \textbf{r}_i - \textbf{r}_j$, $r_{ij} = |\textbf{r}_{ij}|$.

If $\delta \gg 1$, the terms of the order of $\mathrm{erfc}(\delta n)/n$ can be omitted since $\lim\limits_{\delta \to \infty} \mathrm{erfc}(\delta n)/n = 0$ for any $n > 0$ (for example,  $\mathrm{erfc}(5)\approx 10^{-12}$). 
Thus, formula \eqref{eq:ewaldexact} simplifies to:
\begin{equation}
\label{eq:EwaldFormula}
E =\phi_1\sum_{i = 1}^NQ_i^2+ \cfrac{1}{2}\,\sum_{i = 1}^N\sum_{\substack{j = 1\\
		i\neq j}}^N Q_iQ_j\phi_2(\textbf{r}_{ij}),
\end{equation}
where
\begin{equation}
\label{eq:unarPot}
\phi_1 = \cfrac{1}{L}\left[\cfrac{1}{2\pi}\sum_{\textbf{n}\neq\textbf{0}}\exp\left(-\cfrac{\pi^2}{\delta^2}\,n^2\right)n^{-2}-\cfrac{\delta}{\sqrt{\pi}}\right],
\end{equation}
\begin{equation}
\label{eq:pairPot}
\phi_2(\textbf{r}_{ij}) = \cfrac{1}{L}\left[
\cfrac{{\rm erfc}(\delta r_{ij}/L)}{r_{ij}/L} 
+ \cfrac{1}{\pi}\sum_{\textbf{n}\neq \textbf{0}}\exp\left(-\cfrac{\pi^2}{\delta^2}\,n^2\right)n^{-2}\cos\left(2\pi \textbf{n}\cdot \textbf{r}_{ij}/L\right)
\right].
\end{equation}
We will call Eqs. \eqref{eq:unarPot}--\eqref{eq:pairPot} the Ewald potential and Eq.~\eqref{eq:EwaldFormula} the Ewald formula.
It is worth to note that formula \eqref{eq:EwaldFormula} contains the summation over $N$ particles only. The interaction with all periodic images of the particles is included into the Ewald potential \eqref{eq:unarPot}--\eqref{eq:pairPot}. Thus, formula~\eqref{eq:EwaldFormula} is consistent with the ``minimum--image convention'' employed in many Monte-Carlo calculations. According to this convention, a particle in the main cell is allowed to interact only with each of the $N-1$ other particles in the main cell or with the nearest image of that particle in one of the neighboring cells. In other words, each particle interacts with the $N-1$ particles that happen to be located in a cube centered at the particle \cite[Sec. III]{Brush:JCP:1966}. We are going to explain this concept in more detail in section~\ref{sec:analysis}.

The unary potential $\phi_1$ does not depend on $\textbf{r}_{ij}$; the pair potential $\phi_2(\textbf{r}_{ij})$ defines the electrostatic interaction and is angular dependent. In disordered and isotropic media, such as electrolyte, ionic liquid or plasma, this dependence is confusing and results in additional complications. Thus, our goal is to somehow make the Ewald potential spherically symmetric.

To do it, we use the approach of E. Yakub and C. Ronchi \cite{Yakub:2003}.
Following them, we average Eq. \eqref{eq:pairPot} over all directions of $\textbf{r}$ at a distance $r_{ij}$, since all spatial orientations are equivalent:
\begin{equation}
\phi^{a}_2(r_{ij}) = \cfrac{1}{4\pi}\int\limits_{-1}^1d(\cos\theta)\int\limits_{0}^{2\pi}\phi_2(\textbf{r}_{ij})d\psi.
\end{equation}
The only factor to average is the cosine ($\textbf{n}\cdot \textbf{r}_{ij} = nr_{ij}\cos\theta$):
\begin{equation}
\cfrac{1}{4\pi}\int\limits_{-1}^1 d(\cos\theta) \int\limits_{0}^{2\pi}\cos\left(2\pi nr_{ij}\cos\theta/L\right)d\psi =
\cfrac{L/r_{ij}}{2\pi n}\,\sin\left(2\pi nr_{ij}/L\right).
\end{equation}
Thus, we get an averaged pair potential $\phi^{a}_2(r_{ij})$:
\begin{equation}
\label{eq:pairPotAv}
\phi^{a}_2(r_{ij}) = \cfrac{1}{r_{ij}}\left[
{\rm erfc}(\delta r_{ij}/L)
+ \cfrac{1}{2\pi^2}\sum_{\textbf{n}\neq \textbf{0}}\exp\left(-\pi^2n^2/\delta^2\right)n^{-3}\sin\left(2\pi nr_{ij}/L\right)
\right].
\end{equation}
We expand it in the converging series of $r_{ij}$, expanding ${\rm erfc}(\delta r_{ij}/L)$ and $\sin(2\pi nr_{ij}/L)$ into the Taylor series:
\begin{equation}
\label{eq:averagePotSeries}
\phi^{a}_2(r_{ij}) = \cfrac{1}{r_{ij}}\left(1+
\sum_{k = 1}^{+\infty}C_k r_{ij}^{2k+1}
\right)
\end{equation}
with the coefficients
\begin{equation}
C_k = \cfrac{2(-1)^k}{(2k+1)L^{2k+1}}\left[\cfrac{(2\pi)^{2k-1}}{(2k)!}\sum_{\textbf{n}\neq\textbf{0}}f_{k}(\textbf{n})
-\cfrac{\delta^{2k+1}}{\sqrt{\pi}k!}\right].
\label{eq:Ck}
\end{equation}
This series converges for any real $r_{ij}$, since the Taylor series for ${\rm erfc}(w)$ and $\sin(w)$ converge for any real argument $w$.
Here, we introduced the notation
\begin{equation}
\label{eq:fdef}
f_{k}(\textbf{n}) = f_{k}(n) = \exp\left(-\pi^2n^2/\delta^2\right)n^{2(k-1)}.
\end{equation}

One can easily find $C_0$:
\begin{equation}
C_0 = \cfrac{2}{L}\left[\cfrac{1}{2\pi}\sum_{\textbf{n}\neq\textbf{0}}\exp\left(-\pi^2n^2/\delta^2\right)n^{-2} - \cfrac{\delta}{\sqrt{\pi}}\right] = 2\phi_1,
\end{equation}
which is two times larger than the unary potential $\phi_1$.
To compute $C_k$ for $k\geq 1$, we need to sum an infinite series over $\textbf{n}$. Further computations are made for $k\geq 1$.

We are going to include the zero term $\textbf{n}=\textbf{0}$ in the sum of Eq.~\eqref{eq:Ck}. Since $f_k(\textbf{0}) = \delta_{1,k}$:
\begin{equation}
\label{eq:SumDefinition}
\sum_{\textbf{n}\neq\textbf{0}}f_{k}(\textbf{n}) = \sum_{\textbf{n}}f_{k}(\textbf{n}) - \delta_{1,k}.
\end{equation}
Here, $\delta_{ij}$ is the Kronecker delta.
Now Eq.~\eqref{eq:Ck} transforms into the following expression:
\begin{equation}
\label{eq:coefsSeries}
C_k = \cfrac{2(-1)^k}{(2k+1)L^{2k+1}}\left[\cfrac{(2\pi)^{2k-1}}{(2k)!}\sum_{\textbf{n}}f_{k}(\textbf{n})
-\cfrac{\delta^{2k+1}}{\sqrt{\pi}k!}\right]
+ \cfrac{2\pi}{3L^{3}}\,\delta_{1,k}
\end{equation}
for $k\geq 1$. Thus, we need to exactly calculate the following series over all integer vectors $\textbf{n}$:
\begin{equation}
\label{eq:sum}
\sum_{\textbf{n}}f_{k}(\textbf{n}), \quad k\geq 1.
\end{equation}
Below we compute such a series for any $\delta > 0$. In \cite{Yakub:2003} it is stated, that $C_k = 0$ for $k\geq 2$. Below we demonstrate that this is correct in the limit $\delta\to\infty$. Thus, the most interesting case of $\delta \to \infty$ will be considered separately (see Sec. \ref{sec:limitDeltaInf}). Also, we formally investigate the case $\delta\to 0$ (see Sec. \ref{subsubsec:limit0}). We provide numerical computations of series \eqref{eq:sum} at different values of $\delta$ for $1\leq k \leq 4$ in App. \ref{ap:seriesk14}. The summation of \eqref{eq:sum} is the key result of our work: it proves that E. Yakub's and C. Ronchi's results \cite[Eqs. (6)-(8)]{Yakub:2003} are correct in the limit $\delta \to \infty$.

\section{Summation}
\label{sec:summation}
We consider two ways to calculate series \eqref{eq:sum} using the Euler--Maclaurin and Poisson summation formulas. Both of these formulas transform the sought series into an integral form with some residual terms. Next, we formulate these formulas in case of three dimensions as theorems.
\subsection{Formulation of summation formulas}
There are more general formulations and relations for the Euler--Maclaurin theorem. They can be found in \cite[Sec. 3, Eq. (5)]{Muller:1980}, \cite[Eq. (9)]{Ivanov:1963}, \cite{Pogany:2005}. We use a more practical and simple form of the formula. In the following, $D$ is a regular region, i.e., a region with boundary $\partial D$ for which Green's integral theorem is valid \cite{Muller:1980}.
\begin{theorem}[Euler--Maclaurin]
	\label{th:euler}
	Let $D \subset \mathbb{R}^3$ be a regular region with continuously differentiable boundary surface $\partial D$. Let $f(\textbf{r}): \mathbb{R}^3\to \mathbb{R}$ be a twice continuously differentiable function in $\bar{D} = D\cup \partial D$ and let $\textbf{h}$ be the unit outward
	normal to $\partial D$. Then
	\begin{equation}
	\label{eq:sumEulerFormula}
	\sum_{\textbf{n}\in D} f(\textbf{n}) = \int\limits_D f(\textbf{r})d^3r+\int\limits_{\partial D}\bigl(G(\textbf{r})\nabla f(\textbf{r})-f(\textbf{r})\nabla G(\textbf{r})\bigr)\cdot\textbf{h}ds-\int\limits_D G(\textbf{r}) \Delta f(\textbf{r})d^3r,
	\end{equation}
	where
	\begin{equation}
	G(\textbf{r}) = \cfrac{1}{4\pi^2}\sum_{\textbf{q}\neq\textbf{0}}\cfrac{e^{i2\pi \textbf{q}\cdot \textbf{r}}}{q^2}.
	\end{equation}
	Here, $\textbf{q} = (q_x, q_y, q_z)$ is an integer vector ($q_x, q_y, q_z \in \mathbb{Z}$), $q^2 = q_x^2+q_y^2+q_z^2$, $\textbf{q}\cdot \textbf{r} = q_xr_x + q_yr_y + q_zr_z$. The summation $\sum_{\textbf{q}\neq\textbf{0}}$ means that the term $\textbf{q}=(0,0,0)$ is omitted.
\end{theorem}
The Poisson formula \cite[Theorem 6.11]{Sawano:2011} imposes stronger conditions on the function $f(\textbf{r})$.
\begin{theorem}[Poisson]
	Let $f(\textbf{r}): \mathbb{R}^3\to \mathbb{R}$ be a Schwartz function. Then
	\begin{equation}
	\label{eq:poissonsumformula}
	\sum_{\textbf{n}\in \mathbb{Z}^3}f(\textbf{n}) = \sum_{\textbf{q}\in \mathbb{Z}^3}F(\textbf{q}),
	\end{equation}
	where
	\begin{equation}
	\label{eq:fourierTransformOrig}
	F(\textbf{q})= \int\limits_{\mathbb{R}^3} e^{-2\pi i\textbf{q}\cdot \textbf{r}} f(\textbf{r})d^3r
	\end{equation}
	is a Fourier transform of $f(\textbf{r})$.
\end{theorem}
\noindent The definition of a Schwartz function can be found in \cite[Definition 5.1.]{Sawano:2011}.

Both (\ref{eq:sumEulerFormula}) and (\ref{eq:poissonsumformula}) should give the same result in practice. In App. \ref{app:EulerTopoiison} we give an idea how the Euler--Maclaurin formula \eqref{eq:sumEulerFormula} can be \emph{formally} reduced to the Poisson formula \eqref{eq:poissonsumformula}. 

It is easier to sum series \eqref{eq:sum} using the Poisson formula \eqref{eq:poissonsumformula}, that we are going to perform in the next subsection. Nevertheless, in App.~\ref{app:eulermac} we provide the summation of \eqref{eq:sum} using the Euler--Maclaurin formula.

\subsection{Summation using the Poisson formula}
To calculate \eqref{eq:sum}, we use the Poisson summation formula \eqref{eq:poissonsumformula}:
\begin{equation}
\label{eq:fourier}
\sum_{\textbf{n}}f_{k}(\textbf{n}) = \sum_{\textbf{q}}F_{k}(\textbf{q}),
\end{equation}
where
\begin{equation}
\label{eq:fourierTransform}
F_{k}(\textbf{q})= \int e^{-2\pi i\textbf{q}\cdot\textbf{n}} f_{k}(\textbf{n})d^3{n} = 
\cfrac{2\delta^{2k+1}}{\pi^{2k}}\,\Gamma(k+1/2)e^{-\delta^2q^2}M(1-k,3/2,\delta^2q^2)
\end{equation}
is a Fourier transform of $f_{k}(\textbf{n})$. The summation is now performed over an integer vector $\textbf{q}~=~(q_x, q_y, q_z)$, $q_x, q_y, q_z~\in~\mathbb{Z}$. $M(a, b, x)$ is the confluent hypergeometric function defined by the series:
\begin{equation}
\label{eq:hypergeometricDef}
M(a, b, x) = 
\sum _{{s=1}}^{\infty }{\frac{a^{{(s)}}}{b^{{(s)}}s!}}\,x^{s},
\end{equation}
where $a^{{(s)}}$ denotes the rising factorial:
\begin{equation}
a^{(0)}=1, a^{{(s)}}=a(a+1)(a+2)\cdots (a+s - 1).
\end{equation}

\subsubsection{General formula}
By definition \eqref{eq:hypergeometricDef}:
\begin{equation}
M(1-k,3/2,x) =\sum_{{s=0}}^{\infty }{\frac  {(1-k)^{{(s)}}}{(3/2)^{{(s)}}s!}}\,x^{s}.
\end{equation}
Since $(1-k)^{{(k)}} = 0$,  $(1-k)^{{(k+1)}} = 0$ and so on, the series is truncated:
\begin{equation}
\label{eq:hyperFuncres}
M(1-k,3/2,x) =\sum_{{s=0}}^{k - 1}a_{s, k}x^{s}, \quad a_{s, k} = \frac  {(1-k)^{{(s)}}}{(3/2)^{{(s)}}s!},\quad a_{0, k} = 1.
\end{equation}
Substituting \eqref{eq:hyperFuncres} to \eqref{eq:fourierTransform}, we get series \eqref{eq:sum} in the following form:
\begin{equation}
\label{eq:sumsumsum}
\sum_{\textbf{n}}f_{k}(\textbf{n}) = \cfrac{2\delta^{2k+1}}{\pi^{2k}}\,\Gamma(k+1/2)
\sum_{{s=0}}^{k - 1}a_{s, k}\delta^{2s}
\sum_{\textbf{q}}e^{-\delta^2q^2}q^{2s}.
\end{equation}
To perform the summation over $\textbf{q}$, we use the multinomial theorem:
\begin{multline}
	\label{eq:qseries}
	\sum_{\textbf{q}}q^{2s}e^{-\delta^2q^2} = \sum_{\textbf{q}}(q_x^2+q_y^2+q_z^2)^{s}e^{-\delta^2q^2} = \sum_{\textbf{q}}
	\sum\limits_{\alpha_1+\alpha_2+\alpha_3=s}
	\cfrac{s!}{\alpha_1!\alpha_2!\alpha_3!}\,q_x^{2\alpha_1}q_y^{2\alpha_2}q_z^{2\alpha_3}e^{-\delta^2q_x^2}e^{-\delta^2q_y^2}e^{-\delta^2q_z^2} = \\
	=
	\sum\limits_{\alpha_1+\alpha_2+\alpha_3=s}\cfrac{s!}{\alpha_1!\alpha_2!\alpha_3!}
	\left(\sum_{q_x = -\infty}^{+\infty}q_x^{2\alpha_1}e^{-\delta^2q_x^2}\right)
	\times
	\left(\sum_{q_y=-\infty}^{+\infty}q_y^{2\alpha_2}e^{-\delta^2q_y^2}\right)
	\times
	\left(\sum_{q_z=-\infty}^{+\infty}q_z^{2\alpha_3}e^{-\delta^2q_z^2}\right).
\end{multline}
The summation over $\alpha_1, \alpha_2, \alpha_3 \geq 0$ is performed only if $\alpha_1+\alpha_2+\alpha_3=s$.
In this way, we separated the variables so that all sums became one-dimensional. Each internal sum is related with a Jacobi theta function with zero argument:
\begin{equation}
\sum_{q = -\infty}^{+\infty}q^{2\alpha}e^{-\delta^2q^2} =
(-1)^{\alpha}\left(\cfrac{1}{2\delta}\cfrac{\partial }{\partial \delta}\right)^{\alpha}\vartheta _3(0,e^{-\delta^2}),
\end{equation}
where $\vartheta _3(0,x)$ is defined by:
\begin{equation}
\vartheta _3(0,x) = \sum_{q = -\infty}^{+\infty}x^{q^2} = 1 + 2\sum_{q = 1}^{+\infty}x^{q^2}, \quad |x| < 1.
\end{equation}
Thus, we get the final formula substituting \eqref{eq:qseries} to \eqref{eq:sumsumsum}:
\begin{multline}
	\label{eq:mainResultSum}
	\sum_{\textbf{n}}f_{k}(\textbf{n}) =  
	\cfrac{2}{\pi^{2k}}\,\Gamma(k+1/2)\delta^{2k+1} 
	\sum_{s=0}^{k-1}
	a_{s,k}\delta^{2s} \cfrac{(-1)^{s}s!}{2^{s}}\\\times
	\sum\limits_{\alpha_1+\alpha_2+\alpha_3=s}\cfrac{1}{\alpha_1!\alpha_2!\alpha_3!}
	\left(\cfrac{1}{\delta}\cfrac{\partial }{\partial \delta}\right)^{\alpha_1}\vartheta _3(0,e^{-\delta^2}) \times
	\left(\cfrac{1}{\delta}\cfrac{\partial }{\partial \delta}\right)^{\alpha_2}\vartheta _3(0,e^{-\delta^2}) \times
	\left(\cfrac{1}{\delta}\cfrac{\partial }{\partial \delta}\right)^{\alpha_3}\vartheta _3(0,e^{-\delta^2}).
\end{multline}
So the summation over unrestricted three-dimensional argument is transformed into a finite sum. 

We tested formula \eqref{eq:mainResultSum} for $1\leq k\leq 4$ using Wolfram Mathematica \cite{Mathematica}: numerical summation of $\sum_{\textbf{n}}f_{k}(\textbf{n})$ and symbolic calculation of the right part of \eqref{eq:mainResultSum} for $0.8\leq \delta\leq 3$ gives the same results with a machine accuracy. Here, the limitation $\delta\geq 0.8$ appears because Mathematica fails to compute the final numerical result for small $\delta$. The reasons for this fact is unclear to us and is beyond the scope of this work. We hope, that formula \eqref{eq:mainResultSum} will be useful for numerical calculations of Jacobi theta function $\vartheta _3(0,x)$ derivatives (see App. \ref{ap:seriesk14}).

The most interesting and practical result is produced in the limit $\delta\to \infty$. It corresponds to an infinitely small width of the normally distributed charge.

\subsubsection{The limit of infinitely small width $\sqrt{L^2/(2\delta^2)}$}
\label{sec:limitDeltaInf}
In the limit $\delta\to\infty$, theta function becomes constant:
\begin{equation}
\lim\limits_{\delta\to\infty}\vartheta _3(0,e^{-\delta^2}) = 
\lim\limits_{x\to0}\vartheta _3(0,x) = 1.
\end{equation}
Thus, only zero-order derivatives in \eqref{eq:mainResultSum} gives a non-zero result; therefore the only term at $\alpha_1=\alpha_2=\alpha_3=0 = s$ contributes to \eqref{eq:mainResultSum}. The final result for all $k\geq 1$ is:
\begin{equation}
\label{eq:series_asympt_inf}
\sum_{\textbf{n}}f_{k}(\textbf{n}) = \cfrac{2}{\pi^{2k}}\,\Gamma(k+1/2)\delta^{2k+1},\quad \delta\to\infty.
\end{equation}
This asymptotic behavior is valid even for relatively small values of $\delta$ (see App. \ref{ap:seriesk14}, Fig. \ref{fig:seriesdelta}).
Using $\Gamma(k+1/2)=\frac{(2k)!\sqrt{\pi}}{2^{2k}k!}$, we obtain all the coefficients for $k \geq 1$:
\begin{equation}
C_k = \cfrac{2\pi}{3L^{3}}\,\delta_{1,k},\quad \delta\to\infty.
\end{equation}
This key result was presented in \cite{Yakub:2003} without any proof.

Now the averaged pair potential takes a simple form:
\begin{equation}
\phi^a_2(r_{ij}) = (1+C_0r_{ij}+C_1r_{ij}^3)/r_{ij}.
\end{equation}
The full potential energy $E$ is then replaced with $E^a$:
\begin{equation}
\label{eq:energy1}
E^a =\cfrac{1}{2}\sum_{i = 1}^NQ_i^2C_0+ \cfrac{1}{2}\sum_{i = 1}^N\sum_{\substack{j = 1\\
		i\neq j}}^N \cfrac{Q_iQ_j}{r_{ij}}\,(1+C_0r_{ij}+C_1r_{ij}^3).
\end{equation}
The first constant term in Eq. \eqref{eq:energy1} is eliminated by $C_0$ in the second term due to the electroneutral condition \eqref{eq:electroneutral}:
\begin{equation}
\cfrac{1}{2}\sum_{i = 1}^NQ_i^2C_0+ \cfrac{1}{2}\sum_{i = 1}^N\sum_{\substack{j = 1\\
		i\neq j}}^N \cfrac{Q_iQ_j}{r_{ij}}\,C_0r_{ij} = 
\cfrac{1}{2}\sum_{i = 1}^NQ_i\sum_{j = 1}^NQ_jC_0 = 0.
\end{equation}
Total energy results in
\begin{equation}
E^a = \cfrac{1}{2}\sum_{i = 1}^N\sum_{\substack{j = 1\\
		i\neq j}}^N \cfrac{Q_iQ_j}{r_{ij}}\left(1+\cfrac{2\pi}{3L^{3}}\,r_{ij}^3\right)
=
\cfrac{1}{2}\sum_{i = 1}^N\sum_{\substack{j = 1\\
		i\neq j}}^N Q_iQ_j \varphi(r_{ij}),
\end{equation}
\begin{equation}
\label{eq:averagedPot}
\varphi(r) = \cfrac{1}{r}\left[1+\cfrac{1}{2}\,
\left(\cfrac{r}{r_m}\right)^3\right],
\end{equation}
where $r_m = (\tfrac{3}{4\pi})^{1/3}L$ is the radius of the sphere $\frac{4\pi}{3}r^3_m = L^3$ with equivalent volume $L^3$. We will call Eq. \eqref{eq:averagedPot} the averaged potential or the angular--averaged Ewald potential.

Eq.~\eqref{eq:averagedPot} can be used in practical calculations of two-component plasma energy and is independent of any external parameters. In Sec. \ref{sec:analysis}, we describe the main properties of the averaged potential.

Despite Eqs. \eqref{eq:EwaldFormula}, \eqref{eq:averagePotSeries} are valid only if $\delta \gg 1$, we can formally investigate the behavior of series \eqref{eq:mainResultSum} in the limit $\delta\to 0$. 

\subsubsection{The limit $\delta \to 0$}
\label{subsubsec:limit0}
In the limit $\delta\to0$, theta function shows hyperbolic behavior:
\begin{equation}
\vartheta _3(0,e^{-\delta^2}) =  \cfrac{\sqrt{\pi}}{\delta}\, ,\quad \delta\to 0.
\end{equation}
First, we calculate the derivative of theta function in the limit $\delta \to 0$:
\begin{equation}
\left(\cfrac{1}{2\delta}\cfrac{\partial }{\partial \delta}\right)^{\alpha}\vartheta _3(0,e^{-\delta^2})= 
\sqrt{\pi}\left.\left(\cfrac{\partial }{\partial x}\right)^{\alpha}\cfrac{1}{\sqrt{x}}\,\right|_{x=\delta^2} = 
\cfrac{(2\alpha)!\sqrt{\pi}}{\delta^{2\alpha+1}(-1)^\alpha 4^\alpha\alpha!}\, ,\quad \delta\to 0.
\end{equation}
The sought \eqref{eq:mainResultSum} then become:
\begin{equation}
\sum_{\textbf{n}}f_{k}(\textbf{n}) = 
\cfrac{2}{\pi^{2k}}\,\Gamma(k+1/2)\delta^{2k-2} \pi\sqrt{\pi}
\sum_{s=0}^{k-1}
a_{s,k}\, \cfrac{s!}{2^{2s}}\,
\sum\limits_{\alpha_1+\alpha_2+\alpha_3=s}\cfrac{(2\alpha_1)!(2\alpha_2)!(2\alpha_3)!}{(\alpha_1!\alpha_2!\alpha_3!)^2}\, ,\quad \delta\to 0.
\end{equation}
We rewrite the sum over $\alpha_1, \alpha_2, \alpha_3$ as follows
\begin{equation}
A(s) = \sum\limits_{\alpha_1+\alpha_2+\alpha_3=s}\cfrac{(2\alpha_1)!(2\alpha_2)!(2\alpha_3)!}{(\alpha_1!\alpha_2!\alpha_3!)^2}
=
\sum_{\alpha_1=0}^{s}\sum_{\alpha_2=0}^{s-\alpha_1}\cfrac{(2\alpha_1)!(2\alpha_2)!(2(s-\alpha_1-\alpha_2))!}{(\alpha_1!\alpha_2!(s-\alpha_1-\alpha_2)!)^2},
\end{equation}
and introduce the notation:
\begin{equation}
S(k) = \sum_{s=0}^{k-1}
a_{s,k} \cfrac{s!}{2^{2s}}\,A(s).
\end{equation}
One can easily find that $S(1) = 1$. We could not derive an explicit expression for $S(k)$ for any $k\geq 2$; nevertheless exact symbolic calculations using Mathematica \cite{Mathematica} give $S(k) = 0$ for any $2\leq k\leq 60$. We suppose, that
\begin{equation}
S(k) = \delta_{1,k},
\end{equation}
that results in
\begin{equation}
\label{eq:series_asympt_zero}
\sum_{\textbf{n}}f_{k}(\textbf{n})  = \delta_{1,k}, \delta \to 0.
\end{equation}
One can see the dependence of series (\ref{eq:sum}) on $\delta$ for $\delta\ll 1$ in Fig. \ref{fig:seriesdelta}.

\section{Analysis of averaged potential}
\label{sec:analysis}
\begin{figure}[H]
	\centering
	\includegraphics[width=0.6\linewidth]{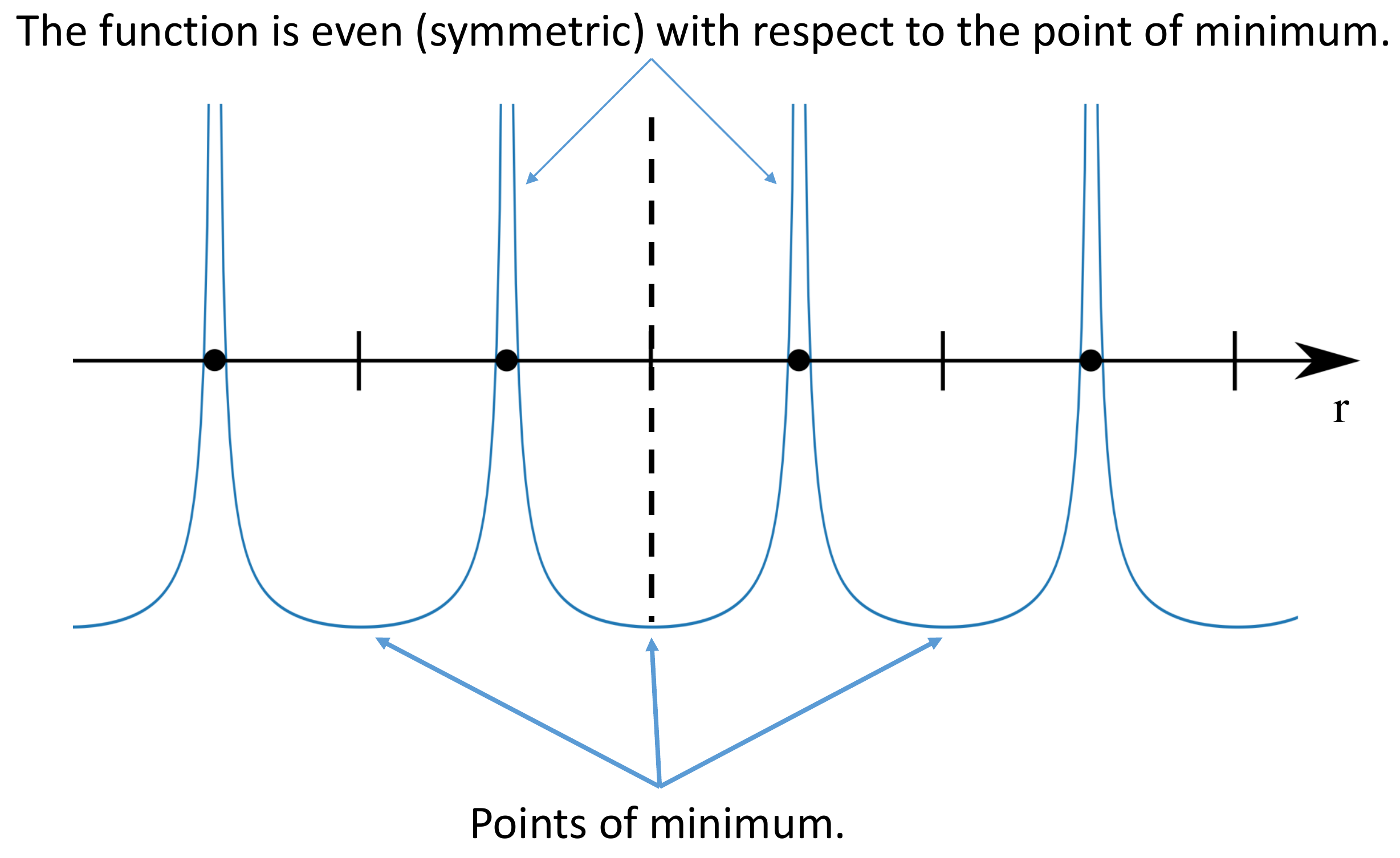}%
	\caption{Qualitative behavior of the Ewald potential along direction [100]. Black dots illustrate the ion positions; vertical lines represent the edges of a unit cell. The solid line is a potential energy. It has a maximum at ion positions and a minimum between ions at the cell edge.}
	\label{fig:periodicpot}
\end{figure}

\begin{figure}[h]
\centering
\includegraphics[width=0.5\linewidth]{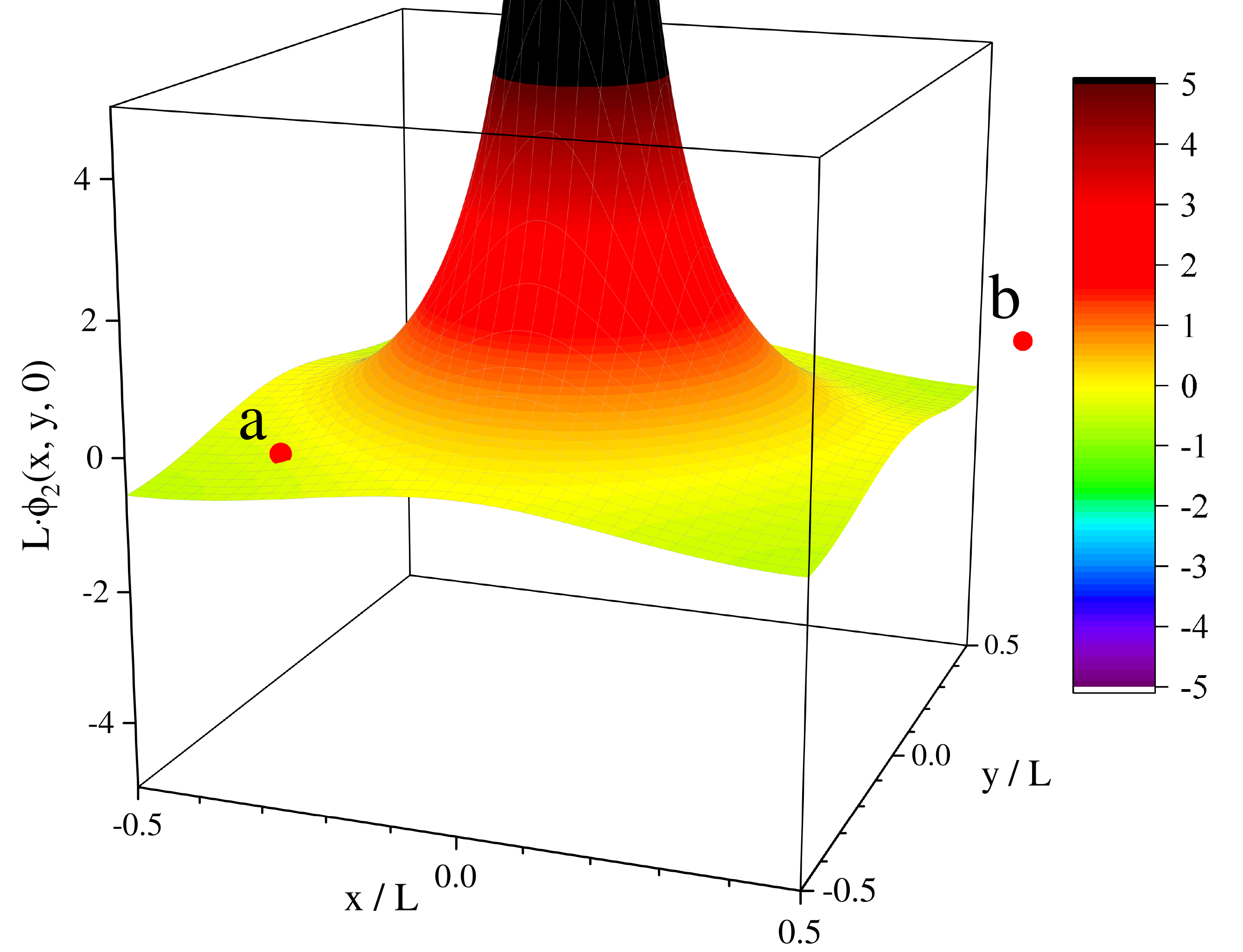}
\caption{A representation of the pair Ewald potential, $L\phi_2(\textbf{r})$. The $i$-th particle is placed in the cube center. If particle ``$a$'' is located inside the cube, it contributes to $u(\textbf{r}_i)$. If particle ``$b$'' is outside the cube, its contribution equals to zero.
So $L\phi_2(\textbf{r})$ can be set to zero if $\mathbf{r}$ is outside the cube.
Thus, the Ewald potential can be considered as short--range.
}
\label{fig:ewaldpotcube}
\end{figure}

Remember that our system has periodic boundary conditions. Therefore, the interaction potential must have the following properties along some chosen direction:
\begin{itemize}
	\item be periodic;
	\item has a minimum at some point;
	\item be symmetrical (even) relative to the point of minimum.
\end{itemize} 
To illustrate this, let us consider the Ewald potential along direction [100] (see Fig.~\ref{fig:periodicpot}). Along other directions, the potential behaves similarly (see Figs.~\ref{fig:ewaldpotcube}, \ref{fig:potentials}). The solid line in Fig.~\ref{fig:periodicpot} shows the potential energy interaction between some trial unit charge at a position $r$ and the system of ions located in the centers of periodically repeated cells (black dots in Fig.~\ref{fig:periodicpot}).
First, if the position of a trial charge is the same as one of the ions, the energy should be infinite. Second, if the position of a trial charge is equidistant from the two ions (at the cell edge), the energy should take a minimum value. These considerations show why an interaction potential has the properties described above.

As we see, the derived potential \eqref{eq:averagedPot} reaches its minimum value at the point $r = r_m$ and then increases infinitely. A common practice \cite{Yakub:2005,Yakub:JPA:2006,Jha:2010,Filinov:PRE:2020, Yakub:2007, Fukuda:2011, Fukuda:2012, Guerrero:2011, Fukuda:2013, Guo:2011, Lytle:2016, Nikitin:2020, KAMIYA201326} (see also the original work \cite{Yakub:2003}), is to consider the expression \eqref{eq:averagedPot} up to $r = r_m$; for $r> r_m$ the potential is redefined by zero. So the averaged potential is truncated at $r= r_m$. We offer the following \emph{qualitative} reasoning that explains such a truncation.

For this purpose, we refer to the calculation procedure using the Ewald formula \eqref{eq:EwaldFormula}. Let us calculate the potential $u(\textbf{r}_i)$ of the $i$-th particle with the coordinate $\textbf{r}_i$:
\begin{equation}
u(\textbf{r}_i) = Q_i \phi_1 + \sum_{\substack{j = 1\\i < j}}^N Q_j\phi_2(\textbf{r}_{ij}).
\end{equation}
The potential $u(\textbf{r}_i)$ is created by the particles around the $i$-th one. Let us surround the $i$-th particle with a cube with side $L$ so that the point $\textbf{r}_i$ is in the center of the cube. To calculate $u(\textbf{r}_i)$, one takes into account only the particles (or their periodic images) inside the cube. 
It means that one can redefine \eqref{eq:pairPot} by zero if $\mathbf{r}_{ij} = \mathbf{r}_j - \mathbf{r}_i$ is outside the cube; this operation has no effect on the value of $u(\textbf{r}_i)$.
The Ewald potential thus modified becomes \emph{short-range} (see Fig. \ref{fig:ewaldpotcube}). 

The range of interaction depends on the direction and is given by the cube surface.
It is interesting that the Ewald potential reaches its first minimum values on the surface of the cube. This means that the partial derivatives, $\partial \phi_2(\textbf{r})/\partial \alpha$, where $\alpha = x, y, z$, are zero on the cube surface, at $\alpha = \pm L/2$. Thus, the first minimum positions of the potential determine its interaction range (see Fig. \ref{fig:ewaldpotcube}).

The averaged potential \eqref{eq:averagedPot} reaches a minimum value $3/(2r_{m}) > 0$ at a point $r=r_m$. Its minimum value, i.e., the range of interaction, is independent of the direction; the points of the potential minimum form the surface of a sphere. The volume of this sphere is $4\pi r_m^3 /3 = L^3$. 

Now we surround the $i$-th particle with a sphere with radius $r_m$ so that $\mathbf{r}_i$ is in the center of the sphere, and calculate $u(\textbf{r}_i)$ using Eq.~\eqref{eq:averagedPot} instead of Eqs.~\eqref{eq:unarPot}--\eqref{eq:pairPot}.
In this case, all particles in the \emph{sphere} of volume $L^3$ must be taken into account.
Therefore, we consider expression \eqref{eq:averagedPot} only up to a distance $r_m$; for $r>r_m$ we redefine $\varphi(r)$ by zero: $\varphi(r) = 0$. This redefinition has no effect on the total potential energy (since the interaction range is $r_m$), but is helpful for the implementation of the calculation algorithm (see Sec. \ref{sec:compAlg}).

\begin{figure}[h!]
	\centering{
\includegraphics[width=0.4\linewidth]{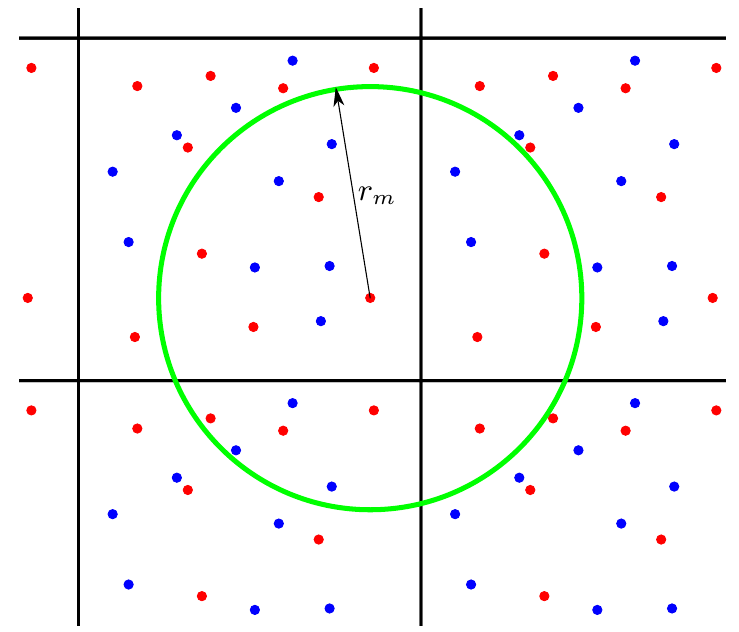}%
		\caption{Example of a two-dimensional system of ions (red and blue dots). The blue ones are charged negatively and the red ones positively. The chosen positive ion interacts only with ions within the green circle with radius $r_m$. The effective interaction with other ions is zero.}
		\label{fig:ion_picture}
	}
\end{figure}

\begin{figure}[H]
	\centering
	\includegraphics[width=0.8\linewidth]{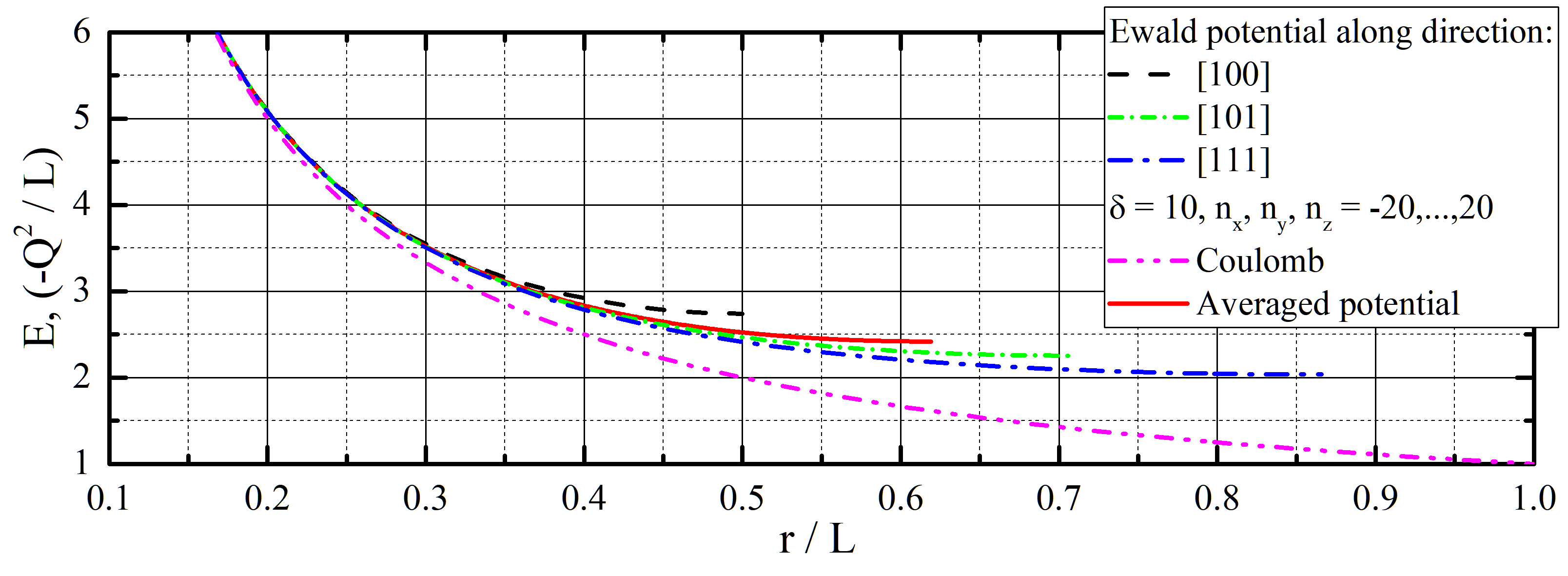}
	\caption{Potential energy (in $-Q^2/L$ units) of two particles with charges $Q_1=-Q_2 = Q$ as a function of reduced distance $r/L$.}
	\label{fig:potentials}
\end{figure}

Then to calculate the potential energy via Eq. \eqref{eq:averagedPot} at the point $\textbf{r}_i$, one has to use the following algorithm:
\begin{enumerate}
	\item Move to the reference point $\textbf{r}_i$ of the selected ion $i$;
	\item Calculate the energy of its interaction with each $j$-th ion, if $|\textbf{r}_i- \textbf{r}_j|\leq r_m$.
\end{enumerate} 
Thus, we proceed to consider a spherical cell, which we superimpose on a periodic cubic cell (see Fig. \ref{fig:ion_picture}). Now the selected ion is affected not only by particles in the main cell but also particles in the sphere. The total number of particles in the sphere $N_s$ depends on the sphere center position.

Now examine what charge density is created by each particle in the sphere.
Consider just one particle in a sphere at the position $\textbf{r}_1$. It creates, at some point $\textbf{r}$, a potential:
\begin{equation}
U(\textbf{r}) = Q_1\varphi(|\textbf{r}-\textbf{r}_1|).
\end{equation}
We calculate the charge density $\rho(\textbf{r})$ at a point $\textbf{r}$:
\begin{equation}
\Delta U(\textbf{r}) = -4\pi\rho(\textbf{r}),
\end{equation}
using the Poisson equation. 
The Laplacian of the averaged potential has the following form:
\begin{equation}
\Delta \varphi(r) = -4\pi\delta(\textbf{r}) + \cfrac{3}{r_m^3}\,.
\end{equation}
Then the charge density:
\begin{equation}
\rho(\textbf{r}) = -\cfrac{\Delta U(\textbf{r})}{4\pi} = Q_1\delta(\textbf{r}-\textbf{r}_1) - \cfrac{3Q_1}{4\pi r_m^3}\,.
\label{eq:charge_density}
\end{equation}
We see that this point particle is not a Coulomb one in the usual sense. 
In addition to the point density $Q_1\delta(\textbf{r}-\textbf{r}_1)$, it creates a uniformly distributed charge of the opposite sign in the entire sphere; its magnitude is  $\tfrac{3Q_1}{4\pi r_m^3}$. This particle can be treated as an ordinary Coulomb point particle + some additional charge around it. The interaction of this additional charge with some other particle is determined by an additional cubic term $\tfrac{1}{2}\,(r/r_m)^3$ in the averaged potential \eqref{eq:averagedPot}.
Moreover, the charge density \eqref{eq:charge_density} is such that the entire sphere is electrically neutral:
\begin{equation}
\iiint\rho(\textbf{r})d^3r = Q_1 - \cfrac{3Q_1}{4\pi r_m^3} \cfrac{4\pi r_m^3}{3} = 0.
\end{equation}
Thus, the averaged potential \eqref{eq:averagedPot} describes the interaction of spheres with radius $r_m$ and zero charge; the spheres interact with each other only if the distance between their centers is less than $r_m$.

Finally, we compare the averaged potential with the Ewald potential along the three primary crystal directions (see Fig. \ref{fig:potentials}) and with the pure Coulomb potential. We picture the Ewald potentials  up to a minimal value since further behavior is trivial (see Fig. \ref{fig:periodicpot} and the reasoning above). The averaged potential is plotted for $r\leq r_m$.
All of these potentials tend to the Coulomb one at small distances, in particular, the averaged one:
\begin{equation}
\lim\limits_{r\to 0}\cfrac{\varphi(r)}{1/r} = 1.
\end{equation}
But for large distances they behave differently. The curve of the averaged potential is situated between the curves of the Ewald potentials along [100] and [111] directions. Another difference is the position of the minimum. The smallest ($r/L = 1/2$) and the largest ($\sqrt{3}/2$) positions are for the Ewald potential along the directions [100] and [111], respectively. The value $r_m/L\approx 0.62$ for $\varphi(r)$ is between them. The fact that $r_m/L > 1/2$, which makes difficulties during the numerical calculations, will be discussed further (see Sec. \ref{sec:compAlg}).

Now we will show some applications and advantages of the obtained potential in practice calculations. 

\section{Applications}
\label{sec:appl}
\subsection{Computation algorithm}
\label{sec:compAlg}
Since the averaged potential is truncated at $r=r_m$, there is a discontinuity in the energy at $r\geq r_m$. It can lead to problems during numerical calculations and simulations \cite[p. 302]{Julian:2003}. To avoid this, we shift the averaged potential to make it zero at $r \ge r_m$:
\begin{equation}
\label{eq:energy2}
E^a = \cfrac{1}{2}\sum_{i = 1}^N\sum_{\substack{j = 1\\
		i\neq j}}^N Q_iQ_j [\varphi(r_{ij}) - \varphi(r_{m}) + \varphi(r_{m})]
=
\cfrac{1}{2}\sum_{i = 1}^N\sum_{\substack{j = 1\\
		i\neq j}}^N Q_iQ_j \tilde{\varphi}(r_{ij}) + \cfrac{1}{2}\sum_{i = 1}^N\sum_{\substack{j = 1\\
		i\neq j}}^N Q_iQ_j\varphi(r_{m}),
\end{equation}
where
\begin{equation}
\tilde{\varphi}(r) = 
\begin{cases}
\cfrac{1}{r}\left[1+\cfrac{1}{2}\,(r/r_m)\left((r/r_m)^2-3\right)\right],  &r\leq r_m\\
0,& r > r_m.
\end{cases}
\end{equation}
The potential $\tilde{\varphi}(r)$ will be used in numerical calculations. We rewrite the last term in \eqref{eq:energy2} in a more simple form due to the electroneutrality condition \eqref{eq:electroneutral}:
\begin{equation}
\cfrac{1}{2}\sum_{i = 1}^N\sum_{\substack{j = 1\\i\neq j}}^N Q_iQ_j\varphi(r_{m})
=
\cfrac{1}{2}\sum_{i = 1}^N\sum_{j = 1}^N Q_iQ_j (1-\delta_{ij})\varphi(r_{m})
=
-\sum_{i = 1}^N\cfrac{3Q_i^2}{4r_{m}}\,.
\end{equation}
Thus, we have the following formula for energy \cite[Eqs. (7), (8)]{Yakub:2003}:
\begin{equation}
E^a = -\sum_{i = 1}^N\cfrac{3Q_i^2}{4r_{m}} + \cfrac{1}{2}\sum_{i = 1}^N\sum_{\substack{j = 1\\
		i\neq j}}^N Q_iQ_j \tilde{\varphi}(r_{ij}).
		\label{eq:energy_final}
\end{equation}
The range of $r_{ij}$ in the second sum of Eq.~\eqref{eq:energy_final} is $r_m > L/2$. It sets up another problem: ion $i$ is affected not only by particles in the main cell but also by their images \cite[Sec. III]{Yakub:2003}. Formally, one can apply the same technique, as in traditional atomistic simulations, using the cut-off radius of a potential. So, every particle in the main cubic cell is surrounded by a sphere with radius $r_m$; all interactions of the central particle with other particles and images inside the sphere are summed (see Fig.~\ref{fig:ion_picture}). The details can be found in \cite[see Sec. 3 and Fig. 2]{Jha:2010}.

\subsection{Madelung constant}
In this section, we apply the averaged potential to compute the energy of a two-component system of charges. For simplicity, we consider several ordered systems of stationary charges, as serious problems arise in simulations of pure Coulomb dense systems of moving charges. We show that even for this case our calculations give accurate results. For ordered systems, the spherical cell is not electroneutral (in the sense of Eq. \eqref{eq:electroneutral}). However, if the number of charges increase the total charge $\sum_i Q_i/N$ tends to zero in the spherical cell. 
Thus, the calculation accuracy improves with the number of charges. 

We will examine several ordered systems and calculate their Madelung constants \cite{Yakub:2005}. The full lattice energy $E^a$ is related to the Madelung constant as follows \cite{Kozhberov:PhD:2018}:
\begin{equation}
E^a = N\cfrac{Q^2}{r_0}\,M.
\end{equation}
For the case of the averaged potential, the formula for the Madelung constant $M$ is:
\begin{equation}
\label{eq:MadelungAv}
M = -\cfrac{3Z_i}{2r_{m}/r_0} + r_0\sum_{\substack{j = 1\\
		i\neq j}}^{N_s} Z_j \tilde{\varphi}(r_{ij}),
\end{equation}
where $r_0$ is the nearest neighbor distance and $Z_i = Q_i/e$ is the charge of a particle with respect to the elementary charge.
$M$ is independent of $i$ since all ions of the same sort are in equivalent positions. The results are presented in Secs. \ref{sec:NaCl}, \ref{sec:CsCl}, \ref{sec:CaF2}. All the lengths are given in the units of $a$, the length of a unit cubic cell.

In all cases, increasing the number of ions $N$ leads to a more accurate value of the Madelung constant. We see that the decrease in relative charge $Z/N$ does not necessarily increase the accuracy of $M$, here $Z = \sum_{i=1}^{N_s} Z_i$ and $N_s$ is the total number of particles in the sphere around a chosen ion. Moreover, the absolute value of $Z$ can increase with $N$. Nevertheless, the convergence for $M$ is obviously observed.  

The exact values of Madelung constants were obtained using Eq.~\eqref{eq:MadelungExact}. They coincide with the values given in \cite{Yakub:2005, Mamode:2017, Kozhberov:PhD:2018}.

\subsubsection{NaCl}
\label{sec:NaCl}
The unit cell of NaCl consists of 8 ions. Positions of Na$^+$ and Cl$^-$ ions are shown in Tab.~\ref{tab:NaClIons}. The dependence of the Madelung constant \eqref{eq:MadelungAv} on the number of ions is given in Tab. \ref{tab:MNaCl}. 
 
\begin{table}[H]
	\centering
	\caption{Coordinates of ions in a unit cubic cell with $L/a = 1$ of rock salt, $r_0/a = 1/2$.}
	\begin{tabular}{|c|c|}
		\hline 
		\rule[-1ex]{0pt}{3.5ex}  & Coordinates \\ 
		\hline 
		\rule[-1ex]{0pt}{3.5ex} Na$^+$ & $\left(0, 0, 0\right), \left(\tfrac{1}{2}, \tfrac{1}{2}, 0\right), \left(\tfrac{1}{2}, 0, \tfrac{1}{2}\right), \left(0, \tfrac{1}{2}, \tfrac{1}{2}\right)$ \\ 
		\hline 
		\rule[-1ex]{0pt}{3.5ex} Cl$^-$ & $\left(0, \tfrac{1}{2}, 0\right), \left(\tfrac{1}{2}, 0, 0\right), \left(0, 0, \tfrac{1}{2}\right), \left(\tfrac{1}{2}, \tfrac{1}{2}, \tfrac{1}{2}\right)$ \\ 
		\hline 
	\end{tabular} 
	\label{tab:NaClIons}%
\end{table}%

\begin{table}[H]
	\centering
	\caption{Madelung constant for crystalline NaCl calculated with the averaged potential \eqref{eq:averagedPot} for a different number of ions $N$.}
	\begin{tabular}{|c|c|c|c|c|c|c|}
		\hline
		\hline
		$L/a$ & $N$ & $N_s-N$ & $Z$ & $Z/N$, \% & $|M|$ & Difference, \% \bigstrut\\
		\hline
		1 & 8 & -1 & -5 & -62.5 & 1.52583 & -12.7 \bigstrut\\
		\hline
		3 & 216 & -13 & -29 & -13.4 & 1.73993 & -0.4 \bigstrut\\
		\hline
		5 & 1000 & 21 & 41 & 4.1 & 1.75509 & 0.4 \bigstrut\\
		\hline
		13 & 17576 & -19 & 5 & 0.03 & 1.74618 & -0.08 \bigstrut\\
		\hline
		29 & 195112 & 55 & 55 & 0.03 & 1.74748 & -0.005 \bigstrut\\
		\hline
		62 & 1906620 & -230 & 25 & 0.001 & 1.74762 & 0.003 \bigstrut\\
		\hline
		135 & 19683000 & 1700 & -293 & -0.001 & 1.74755 & -0.0007 \bigstrut\\
		\hline
		\hline
		\multicolumn{5}{r}{Exact:} & \multicolumn{2}{l}{1.74756} \bigstrut[t]\\
	\end{tabular}%
	\label{tab:MNaCl}%
\end{table}%

\subsubsection{CsCl}
\label{sec:CsCl}
The unit cell of CsCl consists  of 16 ions. Positions of Cs$^+$ and Cl$^-$ ions are shown in Tab.~\ref{tab:CsClIons}. The dependence of the Madelung constant \eqref{eq:MadelungAv} on the number of ions is given in Tab.~\ref{tab:MCsCl}.
\begin{table}[H]
	\centering
	\caption{Coordinates of ions in a unit cubic cell with $L/a = 1$ of caesium chloride structure, $r_0/a = \sqrt{3}/4$.}
	\begin{tabular}{|c|c|}
		\hline 
		\rule[-1ex]{0pt}{3.5ex}  & Coordinates \\ 
		\hline 
		\rule[-1ex]{0pt}{3.5ex} Cs$^+$ & $\left(0, 0, 0\right), \left(\tfrac{1}{2}, 0, 0\right), \left(0, \tfrac{1}{2}, 0\right), \left(0, 0, \tfrac{1}{2}\right), 
		\left(\tfrac{1}{2}, \tfrac{1}{2}, 0\right), \left(\tfrac{1}{2}, 0, \tfrac{1}{2}\right), \left(0, \tfrac{1}{2}, \tfrac{1}{2}\right), \left(\tfrac{1}{2}, \tfrac{1}{2}, \tfrac{1}{2}\right)$ \\ 
		\hline 
		\rule[-1ex]{0pt}{3.5ex} Cl$^-$ & $\left(\tfrac{1}{4}, \tfrac{1}{4}, \tfrac{1}{4}\right), \left(\tfrac{3}{4}, \tfrac{1}{4}, \tfrac{1}{4}\right), \left(\tfrac{1}{4}, \tfrac{3}{4}, \tfrac{1}{4}\right), \left(\tfrac{1}{4}, \tfrac{1}{4}, \tfrac{3}{4}\right), \left(\tfrac{1}{4}, \tfrac{3}{4}, \tfrac{3}{4}\right), \left(\tfrac{3}{4}, \tfrac{1}{4}, \tfrac{3}{4}\right), \left(\tfrac{3}{4}, \tfrac{3}{4}, \tfrac{1}{4}\right), \left(\tfrac{3}{4}, \tfrac{3}{4}, \tfrac{3}{4}\right)$ \\ 
		\hline 
	\end{tabular} 
	\label{tab:CsClIons}%
\end{table}%

\begin{table}[H]
	\centering
	\caption{Madelung constant for crystalline CsCl calculated with the averaged potential \eqref{eq:averagedPot} for a different number of ions $N$.}
	\begin{tabular}{|c|c|c|c|c|c|c|}
		\hline
		\hline
		$L/a$ & $N$ & $N_s-N$ & $Z$ & $Z/N$, \% & $|M|$ & Difference, \% \bigstrut\\
		\hline
		1 & 16 & -1 & -1 & -6.3 & 1.75683 & -0.3 \bigstrut\\
		\hline
		2 & 128 & 9 & 25 & 19.5 & 1.81369 & 2.9 \bigstrut\\
		\hline
		5 & 2000 & -11 & 53 & 2.7 & 1.76123 & -0.1 \bigstrut\\
		\hline
		10 & 16000 & 49 & 1 & 0.01 & 1.76421 & 0.09 \bigstrut\\
		\hline
		23 & 194672 & 129 & 241 & 0.12 & 1.76302 & 0.02 \bigstrut\\
		\hline
		50 & 2000000 & -687 & 17 & 0.001 & 1.76262 & -0.003 \bigstrut\\
		\hline
		107 & 19600700 & -931 & 107 & 0.001 & 1.76267 & -0.0002 \bigstrut\\
		\hline
		\hline
		\multicolumn{5}{r}{Exact:} & \multicolumn{2}{l}{1.76267} \bigstrut[t]\\
	\end{tabular}%
	\label{tab:MCsCl}%
\end{table}%

\subsubsection{CaF$_2$}
\label{sec:CaF2}
The unit cell of CaF$_2$ consists  of 12 ions. Positions of Ca$^{2+}$ and F$^-$ ions are shown in Tab.~\ref{tab:CaF2Ions}.  The dependence of the Madelung constant \eqref{eq:MadelungAv} on the number of ions is given in Tab.~\ref{tab:MCaF2}.
\begin{table}[H]
	\centering
	\caption{Coordinates of ions in a unit cubic cell with $L/a = 1$ of fluorite structure, $r_0/a = \sqrt{3}/4$.}
	\begin{tabular}{|c|c|}
		\hline 
		\rule[-1ex]{0pt}{3.5ex}  & Coordinates \\ 
		\hline 
		\rule[-1ex]{0pt}{3.5ex} Ca$^{2+}$ & $\left(0, 0, 0\right), \left(\tfrac{1}{2}, \tfrac{1}{2}, 0\right), \left(\tfrac{1}{2}, 0, \tfrac{1}{2}\right), \left(0, \tfrac{1}{2}, \tfrac{1}{2}\right)$ \\ 
		\hline 
		\rule[-1ex]{0pt}{3.5ex} F$^-$ & $\left(\tfrac{1}{4}, \tfrac{1}{4}, \tfrac{1}{4}\right), \left(\tfrac{3}{4}, \tfrac{1}{4}, \tfrac{1}{4}\right), \left(\tfrac{1}{4}, \tfrac{3}{4}, \tfrac{1}{4}\right), \left(\tfrac{1}{4}, \tfrac{1}{4}, \tfrac{3}{4}\right), \left(\tfrac{1}{4}, \tfrac{3}{4}, \tfrac{3}{4}\right), \left(\tfrac{3}{4}, \tfrac{1}{4}, \tfrac{3}{4}\right), \left(\tfrac{3}{4}, \tfrac{3}{4}, \tfrac{1}{4}\right), \left(\tfrac{3}{4}, \tfrac{3}{4}, \tfrac{3}{4}\right)$ \\ 
		\hline 
	\end{tabular} 
	\label{tab:CaF2Ions}%
\end{table}%

\begin{table}[H]
 	\centering
 	\caption{Madelung constant for crystalline CaF$_2$ calculated with the averaged potential \eqref{eq:averagedPot} for a different number of ions $N$.}
 	\begin{tabular}{|c|c|c|c|c|c|c|}
 		\hline
 		\hline
 		$L/a$ & $N$ & $N_s-N$ & $Z$ & $Z/N$, \% & $|M|$ & Difference, \% \bigstrut\\
 		\hline
 		1 & 12 & -3 & -6 & -50 & 3.07823 & -6.0 \bigstrut\\
 		\hline
 		3 & 324 & -29 & -34 & -10.5 & 3.27549 & -0.02 \bigstrut\\
 		\hline
 		5 & 1500 & -1 & 94 & 6.3 & 3.28118 & 0.2 \bigstrut\\
 		\hline
 		11 & 15972 & 5 & 298 & 1.9 & 3.27692 & 0.02 \bigstrut\\
 		\hline
 		25 & 187500 & -157 & 286 & 0.15 & 3.27574 & -0.01 \bigstrut\\
 		\hline
 		55 & 1996500 & -39 & -486 & -0.024 & 3.27605 & -0.002 \bigstrut\\
 		\hline
 		118 & 19716400 & -2973 & -1122 & -0.006 & 3.27612 & 0.0002 \bigstrut\\
 		\hline
 		\hline
 		\multicolumn{5}{r}{Exact:} & \multicolumn{2}{l}{3.27611} \bigstrut[t]\\
 	\end{tabular}%
 	\label{tab:MCaF2}%
\end{table}%

\subsection{Rate of convergence, scaling and advantages}
\begin{figure}[H]
\centering
\includegraphics[width=0.8\linewidth]{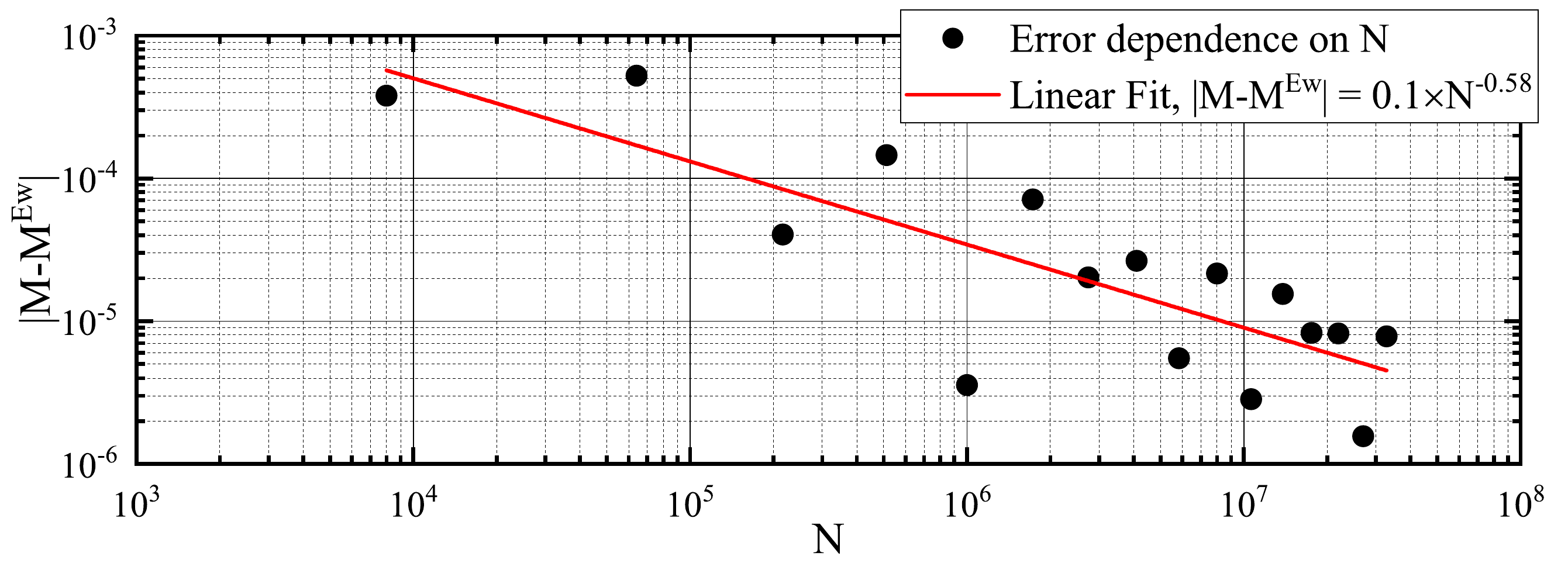}
\caption{Convergence rate of calculations via the averaged potential.}
\label{fig:rateofconvergence}
\end{figure}

In the previous section, we have shown the convergence of the Madelung constant using our computation method for very large ordered systems. Now we are going to estimate the convergence rate of our technique. We calculate the absolute difference of the Madelung constant from the exact value, $|M-M^{\text{Ew}}|$, depending on the number $N$. The calculation for the NaCl lattice gives the results shown in Fig.~\ref{fig:rateofconvergence} on a log--log scale.
A linear approximation of the data in Fig.~\ref{fig:rateofconvergence} was made. Thus $|M-M^{\text{Ew}}| \propto N^{-b}$, where $b = 0.58\pm 0.11$. So with an increase in the number of particles in the cell by $\approx 53$ times, the computational error decreases tenfold. 

One can see that the scatter of the data relative to the fitting line is wide. We explain this by the fact that we apply the method of calculating \emph{a disordered and isotropic} system to \emph{ordered and anisotropic} one. In such a case, one should not expect high accuracy or absence of noise in the data.
Nevertheless, the convergence is observed. 

\begin{figure}[H]
	\centering
	\includegraphics[width=0.8\linewidth]{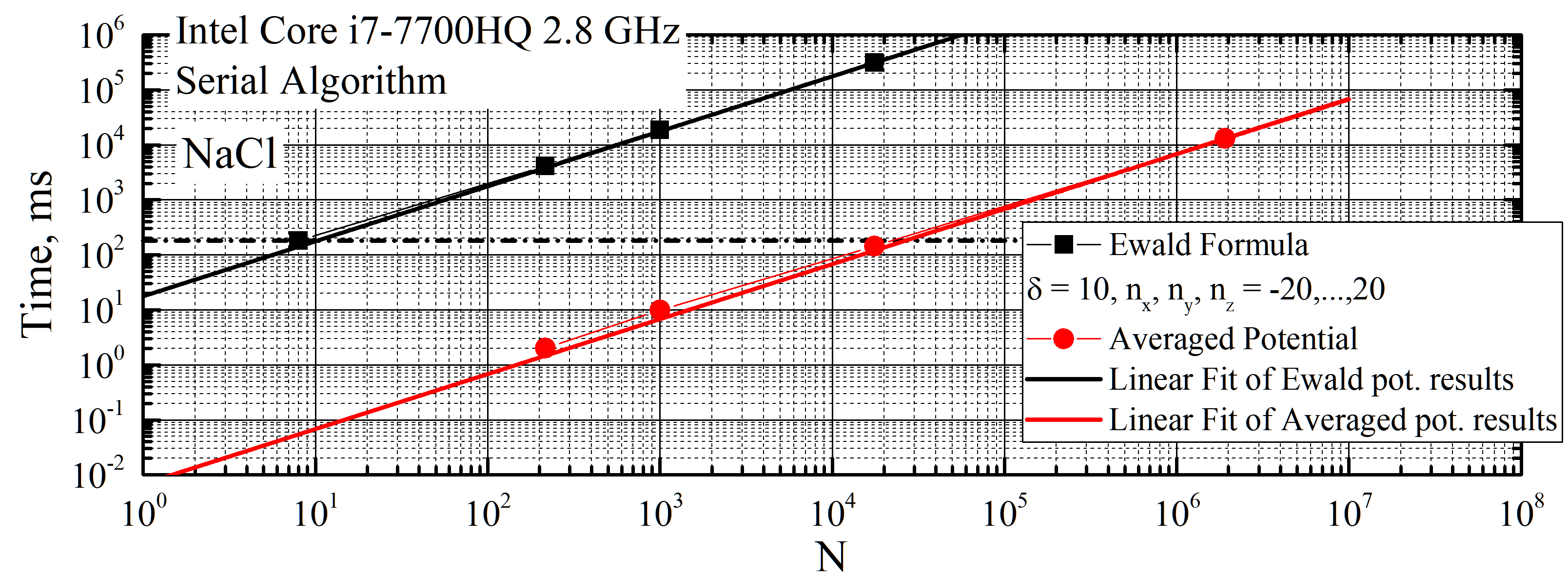}
	\caption{Calculation time of the Madelung constant for NaCl as a function of $N$. Both methods are linear in $N$; the calculation with the averaged potential is 2600 times faster than with the Ewald potential. The slope of the curve obtained with the Ewald potential is $17.7$~ms, obtained with the averaged one is $6.82\times 10^{-3}$~ms.}
	\label{fig:scalinggr}
\end{figure}

It is of practical importance to compare the performance of computations between the exact Ewald formula and averaged potential. The Madelung constant takes the following form with the exact Ewald formula \eqref{eq:ewaldexact} for energy: 
\begin{equation}
\label{eq:MadelungExact}
M^{\text{Ew}} = 2r_0Z_i\phi_1^{\text{Ex}}+r_0\sum_{\substack{j = 1\\i\neq j}}^NZ_j\phi_2^{\text{Ex}}(\textbf{r}_{ij}).
\end{equation}
For comparison, we make calculations using \eqref{eq:MadelungAv} and \eqref{eq:MadelungExact} for NaCl at different $N$. The following parameters for the Ewald potential are chosen: $n_x, n_y, n_z = - 20,\dots, 20$ and $\delta = 10$. We perform sequential calculations on a CPU Intel Core i7-7700HQ 2.8 GHz. Calculation time for different number of ions is present in Fig.~\ref{fig:scalinggr}.
The results were linearly fitted. Although both dependencies are linear, the calculation by the formula \eqref{eq:MadelungAv} is approximately 2600 times faster than by \eqref{eq:MadelungExact}. In \cite[Fig. 3]{Jha:2010} it was also shown that the averaged potential gives faster results compared to the other methods used in paper \cite{Jha:2010}.
 
The difference between the exact result and the one obtained using the averaged potential \eqref{eq:MadelungAv} is about 0.08\% for equivalent computation time (Fig. \ref{fig:scalinggr}, horizontal black line). We hope, that for a disordered system this error will be even smaller.

The averaged potential has a significant advantage: it doesn't depend on any external parameters that affect the convergence. On the other hand, the influence of parameters on the results with the Ewald potential is significant \cite[see Figs. 4-9]{Pratt:2001}.

Thus, the averaged potential is helpful for numerical modeling of Coulomb systems with periodic boundary conditions and a large number of particles in a simulation cell. 
In addition, the averaged potential \eqref{eq:averagedPot} has a fairly simple analytical form compared to the Ewald potential \eqref{eq:pairPot}, that makes it attractive for analytical studies.

Nevertheless, since the averaged potential \emph{is not equivalent} to the Ewald one, it is necessary to check the accuracy of calculation by the proposed method for each Coulomb system. In particular, one should make sure of the convergence on the number of particles before applying the method in practice. 

\section{Conclusion}
\label{Sec:conclusions}
The step by step derivation of the angular--averaged Ewald potential is proposed using the Euler--Maclaurin and Poisson formulas. 
Additionally, the formal equivalence of the Euler--Maclaurin and Poisson formulas is demonstrated. 
From a physical point of view, the averaged potential describes the interaction of two spheres with radius $r_m~=~(3/(4\pi)^{1/3})L$, where $L$ is the size of a cubic computational cell. Each sphere contains a point charge in the center and a compensating uniformly distributed in the entire sphere charge of the same value and opposite sign. The spheres interact with each other only if the distance between their centers is less than $r_m$. Thus, the long--range Coulomb interaction in an disordered point system of charges is replaced with the interaction of electrically neutral spheres with a finite--range potential; the range of the potential depends on the size of a cubic computational cell. Technically, the calculation of the interaction energy is straightforward: every particle in the main cubic cell is surrounded by a sphere with radius $r_m$; all interactions of the central particle with other particles and images inside the sphere are summed. Our computations of the Madelung constant for a number of crystal lattices show the efficiency of the angular--averaged potential for systems containing up to $2\times 10^7$ particles.

\begin{acknowledgments}
The authors thank the Russian Science Foundation (Grant No. 20-42-04421) for financial support. The authors also acknowledge the JIHT RAS
Supercomputer Centre, the Joint Supercomputer Centre of the Russian Academy of Sciences, and the Shared Resource Centre ``Far Eastern Computing Resource'' IACP FEB RAS for providing computing time.
\end{acknowledgments}

\appendix
\section{Series for $1\leq k\leq 4$}
\label{ap:seriesk14}
\begin{figure}[h!]
	\centering
	\includegraphics[width=0.8\linewidth]{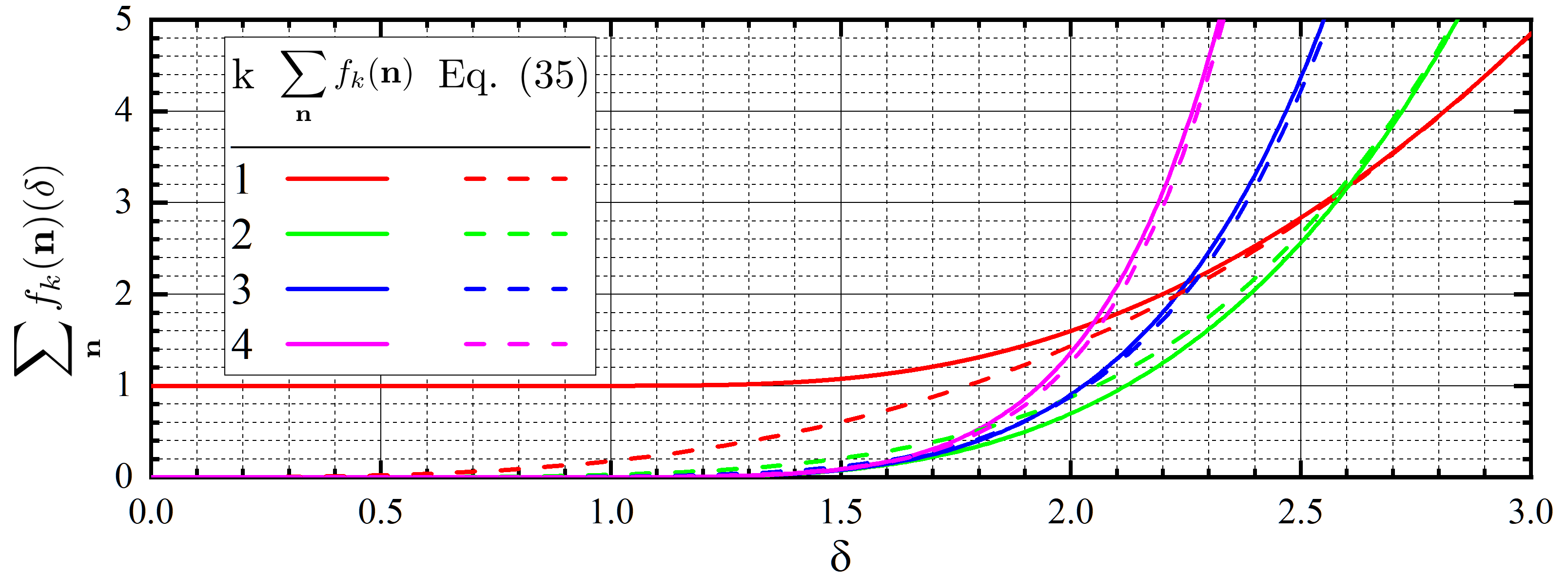}
	\caption{Series \eqref{eq:sum} as a function of $\delta$ for $1\leq k \leq 4$. The solid lines represent numerical summation. The dashed lines are asymptotic behavior at $\delta\to\infty$ \eqref{eq:series_asympt_inf}.}
	\label{fig:seriesdelta}
\end{figure}

Using Eq. \eqref{eq:mainResultSum}, we obtain the direct equations for series $\sum_{\textbf{n}}f_{k}(\textbf{n})$ for $1\leq k\leq 4$:
\begin{equation}
\label{eq:k1series}
\sum_{\textbf{n}}f_{1}(\textbf{n})
=
\frac{\delta^3 \vartheta^3 _3\left(0,e^{-\delta^2}\right)}{\pi ^{3/2}},
\end{equation}
\begin{equation}
\sum_{\textbf{n}}f_{2}(\textbf{n})
=
\frac{3 \delta^5 \vartheta^2 _3\left(0,e^{-\delta^2}\right) \left[\vartheta _3\left(0,e^{-\delta^2}\right)-2 \delta^2 e^{-\delta^2}
	\left.\left(\partial \vartheta_3\left(0,x\right)/\partial x\right)\right|_{x=e^{-\delta^2}}\right]}{2 \pi ^{7/2}},
\end{equation}
\begin{multline}
\sum_{\textbf{n}}f_{3}(\textbf{n})
=
\cfrac{3 \delta^7 e^{-2 \delta^2} \vartheta _3\left(0,e^{-\delta^2}\right)}{4 \pi ^{11/2}}
\left[4 \delta^4 \left(2
	\left.\left(\partial \vartheta_3\left(0,x\right)/\partial x\right)^2\right|_{x=e^{-\delta^2}}+\vartheta _3\left(0,e^{-\delta^2}\right)
	\left.\left(\partial^2 \vartheta_3\left(0,x\right)/\partial x^2\right)\right|_{x=e^{-\delta^2}}\right)
	\right.\\\left.+
	4 \left(\delta^2-5\right) e^{\delta^2} \delta^2 \vartheta _3\left(0,e^{-\delta^2}\right)
	\left.\left(\partial \vartheta_3\left(0,x\right)/\partial x\right)\right|_{x=e^{-\delta^2}}+5 e^{2 \delta^2} \vartheta^2 _3\left(0,e^{-\delta^2}\right)\right],
\end{multline}
\begin{multline}
\label{eq:k4series}
\sum_{\textbf{n}}f_{4}(\textbf{n})
=
-\frac{3 \delta^9 e^{-3 \delta^2}}{8 \pi ^{15/2}} \left[16 \delta^6 \left.\left(\partial \vartheta_3\left(0,x\right)/\partial x\right)^3\right|_{x=e^{-\delta^2}}+
24 \left(2 \delta^2-7\right) e^{\delta^2} \delta^4 \vartheta
_3\left(0,e^{-\delta^2}\right) \left.\left(\partial \vartheta_3\left(0,x\right)/\partial x\right)^2\right|_{x=e^{-\delta^2}}
\right.\\\left.+
2 \delta^2 \vartheta _3\left(0,e^{-\delta^2}\right)
\left.\left(\partial \vartheta_3\left(0,x\right)/\partial x\right)\right|_{x=e^{-\delta^2}} \left(24 \delta^4 \left.\left(\partial^2 \vartheta_3\left(0,x\right)/\partial x^2\right)\right|_{x=e^{-\delta^2}}+\left(4 \delta^4-42
\delta^2+105\right) e^{2 \delta^2} \vartheta _3\left(0,e^{-\delta^2}\right)\right)
\right.\\\left.+
\vartheta^2 _3\left(0,e^{-\delta^2}\right) \left(8 \delta^6
\left.\left(\partial^3 \vartheta_3\left(0,x\right)/\partial x^3\right)\right|_{x=e^{-\delta^2}}
\right.\right.\\\left.\left.
+12 \left(2 \delta^2-7\right) e^{\delta^2} \delta^4
\left.\left(\partial^2 \vartheta_3\left(0,x\right)/\partial x^2\right)\right|_{x=e^{-\delta^2}}-35 e^{3 \delta^2} \vartheta _3\left(0,e^{-\delta^2}\right)\right)\right].
\end{multline}
The dependence of series (\ref{eq:mainResultSum}) on $\delta$ for $1\leq k\leq 4$ are shown in Fig. \ref{fig:seriesdelta} by the solid lines; the dashed lines represent asymptotic behavior at $\delta \to \infty$ \eqref{eq:series_asympt_inf}. The difference between the exact and asymptotic results is small even at $\delta \geq 2.5$; at small $\delta$ the series tends to $0$ for $k\geq 2$ and to $1$ for $k=1$ as it was predicted \eqref{eq:series_asympt_zero}.
We hope, that Eqs. \eqref{eq:mainResultSum}, \eqref{eq:k1series}-\eqref{eq:k4series} will be usefull to numerically calculate derivatives of Jacobi theta function $\vartheta _3(0,x)$.

\section{Idea of transformation of the Euler--Maclaurin to the Poisson summation formula}
\label{app:EulerTopoiison}
In this appendix we want to represent the idea of a relationship between the Euler--Maclaurin and the Poisson summation formulas. Below we present \emph{formal} calculations which show the consequence of the Poisson summation formula from the Euler--Maclaurin one. 
We find this idea interesting and hope that the class of functions $f(\mathbf{r}$ for which our transformations are valid will be defined in the future. 

Our \emph{hypothesis} is the following:
\begin{equation}
	\label{eq:sumEulerFormulaToFourie}
	\sum_{\textbf{n}\in D} f(\textbf{n})  = \sum_{\textbf{q}} \int\limits_D f(\textbf{r})e^{-i2\pi \textbf{q}\cdot \textbf{r}}d^3r,
\end{equation}
where $D \subset \mathbb{R}^3$ is a regular region with continuously differentiable boundary surface $\partial D$ and $f(\textbf{r}): \mathbb{R}^3\to \mathbb{R}$ is a twice continuously differentiable function in $\bar{D} = D\cup \partial D$.

First, we use the following rule for integration by parts \cite[Theorem 37.2]{Rogers:2011}:
\begin{equation}
\int\limits_{D}u(\textbf{r})\nabla\textbf{w}(\textbf{r})d^3r = -\int\limits_{D}\nabla u(\textbf{r})\cdot \textbf{w}(\textbf{r})d^3r + \int\limits_{\partial D}u(\textbf{r})\textbf{w}(\textbf{r})\cdot\textbf{h}ds,
\end{equation}
where $\textbf{h}$ is the unit outward normal to $\partial D$; $u(\textbf{r})$ is a scalar--valued function and \textbf{w}(\textbf{r}) is a vector--valued function and $D \subset \mathbb{R}^3$ is a regular region with continuously differentiable boundary surface $\partial D$.
Since $f(\textbf{r})$ and $D$ satisfies the conditions of Theorem \ref{th:euler}, Eq. \eqref{eq:sumEulerFormula} can be used.
Next, we integrate the last term in \eqref{eq:sumEulerFormula} by parts ($u(\textbf{r}) = G(\textbf{r}), \textbf{w}(\textbf{r}) = \nabla f(\textbf{r})$):
\begin{equation}
\label{eq:proof1}
\int\limits_D G(\textbf{r}) \nabla\cdot\nabla f(\textbf{r})d^3r =-\int\limits_D \nabla G(\textbf{r})\nabla f(\textbf{r})d^3r + 
\int\limits_{\partial D} G(\textbf{r})\nabla f(\textbf{r}) \textbf{h}ds.
\end{equation}
Second, the first Green's identity \cite[Chapter 7, Eq. (G1)]{Walter:2007} will be used:
\begin{equation}
\label{eq:firstGreenId}
\int\limits_{D} \nabla u(\textbf{r})\nabla v(\textbf{r})d^3r = \int\limits_{\partial D}v(\textbf{r})\nabla u(\textbf{r})\cdot\textbf{h}ds - \int\limits_D v(\textbf{r})\Delta u(\textbf{r})d^3r,
\end{equation}
where $v(\textbf{r})$ is a scalar-valued function.
Using (\ref{eq:firstGreenId}), we obtain:
\begin{equation}
\label{eq:proof2}
\int\limits_D \nabla G(\textbf{r})\nabla f(\textbf{r})d^3r = \int\limits_{\partial D} f(\textbf{r})\nabla G(\textbf{r}) \textbf{h}ds - \int\limits_D f(\textbf{r})\Delta G(\textbf{r})d^3r.
\end{equation}
Substituting Eqs. \eqref{eq:proof1}, \eqref{eq:proof2} into the  Euler--Maclaurin summation formula \eqref{eq:sumEulerFormula}, we get:
\begin{equation}
\label{eq:proof3}
\sum_{\textbf{n}\in D} f(\textbf{n}) = \int\limits_D f(\textbf{r})d^3r - \int\limits_D f(\textbf{r})\Delta G(\textbf{r})d^3r.
\end{equation}
It is easy to find $\Delta G(\textbf{r})$:
\begin{equation}
\Delta G(\textbf{r}) = \cfrac{1}{4\pi^2} \sum_{\textbf{q}\neq\textbf{0}}\Delta\cfrac{e^{i2\pi \textbf{q}\cdot \textbf{r}}}{q^2} = -\sum_{\textbf{q}\neq\textbf{0}}e^{i2\pi \textbf{q}\cdot \textbf{r}}.
\end{equation}
Then, the last term in \eqref{eq:proof3} is simplified:
\begin{equation}
\label{eq:proof4}
- \int\limits_D f(\textbf{r})\Delta G(\textbf{r})d^3r =  \sum_{\textbf{q}\neq\textbf{0}} \int\limits_D f(\textbf{r})e^{i2\pi \textbf{q}\cdot \textbf{r}}d^3r = |\textbf{q}\to -\textbf{q}| = \sum_{\textbf{q}\neq\textbf{0}} \int\limits_D f(\textbf{r})e^{-i2\pi \textbf{q}\cdot \textbf{r}}d^3r.
\end{equation}
In the last equality we have inverted the summation order, which is shown by the notation $|\textbf{q}\to -\textbf{q}|$.
Also in Eq.~\eqref{eq:proof4} we have exchanged the limit of the partial sums of the series with the integral \emph{without} any justification.

Now we can include the $\textbf{q} = \textbf{0}$ term into sum \eqref{eq:proof4}:
\begin{equation}
\label{eq:theoremIII2Proof}
\sum_{\textbf{n}\in D} f(\textbf{n}) = \int\limits_D f(\textbf{r})d^3r + \sum_{\textbf{q}\neq\textbf{0}} \int\limits_D f(\textbf{r})e^{-i2\pi \textbf{q}\cdot \textbf{r}}d^3r = \sum_{\textbf{q}} \int\limits_D f(\textbf{r})e^{-i2\pi \textbf{q}\cdot \textbf{r}}d^3r,
\end{equation}
which leads us to the desired relationship \eqref{eq:sumEulerFormulaToFourie}.

Thus, the residual term in the Euler--Maclaurin formula can be written as follows:
	\begin{equation}
	\int\limits_{\partial D}\left(G(\textbf{r})\nabla f(\textbf{r})-f(\textbf{r})\nabla G(\textbf{r})\right)\cdot\textbf{h}ds-\int\limits_D G(\textbf{r}) \Delta f(\textbf{r})d^3r = \sum_{\textbf{q}\neq\textbf{0}} \int\limits_D f(\textbf{r})e^{-i2\pi \textbf{q}\cdot \textbf{r}}d^3r.
	\end{equation}
This form of the residual term in the Euler--Maclaurin formula is more appropriate for practical calculations due to the absence of a surface integral.

If one considers a sphere of radius $R$ as a region $D$ and take the limit $R\to \infty$, formally the Poisson formula \eqref{eq:poissonsumformula} is obtained (see App. \ref{app:eulermac}). However, the class of functions $f(\textbf{r})$, for which such a limit is valid, remains uncertain.

\section{Summation using the Euler--Maclaurin formula}
\label{app:eulermac}
In this appendix we will use the Euler--Maclaurin formula \eqref{eq:sumEulerFormula} to sum series \eqref{eq:sum}.

We will choose a sphere of a radius $R$ as a region $D$ in \eqref{eq:sumEulerFormula}.
Let us first consider the integral over surface:
\begin{equation}
\label{eq:app1}
\int\limits_S\left(f_{k}(\textbf{r})\nabla G(\textbf{r})-G(\textbf{r})\nabla f_{k}(\textbf{r})\right)\textbf{h}ds.
\end{equation}
Since $f_{k}(\textbf{r})$ has an exponent factor, the following term is eliminated if $R\to\infty$:
\begin{equation}
\int\limits_Sf_{k}(\textbf{r})\nabla G(\textbf{r})\textbf{h}ds = \exp\left(-\cfrac{\pi^2}{\delta^2}\,R^2\right)R^{2(k-1)}
\int\limits_S\nabla G(\textbf{r})\textbf{h}ds\xrightarrow{R\to\infty}0.
\end{equation}
The second term in \eqref{eq:app1} is also eliminated, since $\nabla f_{k}(\textbf{r}) \propto \exp\left(-\cfrac{\pi^2}{\delta^2}\,r^2\right)$:
\begin{equation}
\nabla f_{k}(\textbf{r}) = 2r^{2(k-1)}\exp\left(-\cfrac{\pi^2}{\delta^2}\,r^2\right)\left((k-1)r^{-2}-\tfrac{\pi^2}{\delta^2}\right)\textbf{r}.
\end{equation}
Thus, the whole term \eqref{eq:app1} is equal to zero in the limit $R\to\infty$.

Next, we integrate the last term in \eqref{eq:sumEulerFormula}:
\begin{equation}
\int\limits G(\textbf{r}) \Delta f_{k}(\textbf{r})d^3r,
\end{equation}
over $\mathbb{R}^3$, since $R\to\infty$. We introduce the notation:
\begin{equation}
g_k(r) = \Delta f_{k}(r) = 2 e^{-\tfrac{\pi^2}{\delta^2} r^2} r^{-4+2 k} \left(1-3 k+2 k^2+\tfrac{\pi^2}{\delta^2} (1-4 k) r^2+2 \tfrac{\pi^4}{\delta^4} r^4\right).
\end{equation}
We perform the integration in the spherical coordinates:
\begin{equation}
\int\limits G(\textbf{r}) \Delta f_{k}(\textbf{r})d^3r = \cfrac{1}{4\pi^2}\sum_{\textbf{n}\neq\textbf{0}}\int\limits_0^{\infty}r^2dr
\int\limits_0^{2\pi}d\phi
\int\limits_{-1}^1d(\cos\theta)\cfrac{e^{i2\pi qr\cos\theta}}{q^2}\,g_{k}(r),
\end{equation}
where $\theta$ is the angle  between $\textbf{q}$ and \textbf{r}.
Next, we integrate over angles:
\begin{equation}
\int\limits G(\textbf{r}) \Delta f_{k}(\textbf{r})d^3r = \cfrac{1}{2\pi^2}\sum_{\textbf{q}\neq\textbf{0}}q^{-3}\int\limits_0^{\infty}rg_{k}(r)\sin(2\pi qr)dr,
\end{equation}
and over distance $r$:
\begin{multline}
I_k(\textbf{q},\delta) = \cfrac{1}{2\pi^2q^{3}}\,\int\limits_0^{\infty}rg_{k}(r)\sin(2\pi qr)dr \\=
 \frac{2\delta^{2k+1}}{\pi ^{2 k}} 
\cfrac{\Gamma \left(k+1/2\right)}{6k-3}
\left(4 \delta^2 (k-2) q^2 \,
M\left(k-\frac{1}{2},\frac{5}{2},-\delta^2 q^2\right)+\left(6 \delta^2 q^2-6 k+3\right) \, M\left(k-\frac{1}{2},\frac{3}{2},-\delta^2q^2\right)\right),
\end{multline}
where $M(a,b,x)$ is defined by \eqref{eq:hypergeometricDef}.
We include now the term $\textbf{q} = \textbf{0}$ into summation:
\begin{equation}
\int\limits G(\textbf{r}) \Delta f_{k}(\textbf{r})d^3r  = \sum_{\textbf{q}\neq\textbf{0}} I_k(\textbf{q},\delta) = \sum_{\textbf{q}} I_k(\textbf{q},\delta) - I_k(0,\delta),
\end{equation}
where
\begin{equation}
I_k(0,\delta) = -
\frac{2\delta^{2k+1}}{\pi ^{2 k}} 
\Gamma \left(k+1/2\right) = -\int f_k(\textbf{r})d^3r.
\end{equation}
This term eliminate the integral term in \eqref{eq:sumEulerFormula}:
\begin{equation}
\sum_{\textbf{n}} f_k(\textbf{n}) = -\sum_{\textbf{q}} I_k(\textbf{q},\delta)
\end{equation}
Using a symbolic computations by Wolfram Mathematica \cite{Mathematica}, we get:
\begin{equation}
\cfrac{1}{6k-3}
\left(4 \delta^2 (k-2) q^2 \,
M\left(k-\frac{1}{2},\frac{5}{2},-\delta^2 q^2\right)+\left(6 \delta^2 q^2-6 k+3\right) \, M\left(k-\frac{1}{2},\frac{3}{2},-\delta^2q^2\right)\right)  =  -e^{-\delta^2q^2}M(1-k,3/2,\delta^2q^2),
\end{equation}
that leads us to Eqs.~\eqref{eq:fourier}, \eqref{eq:fourierTransform}.

\end{document}